\documentclass[sigconf]{acmart}
\usepackage[most]{tcolorbox}
\usepackage{algorithm,algpseudocode}
\usepackage{array,multirow}
\usepackage{xspace}
\usepackage{pifont}
\usepackage{adjustbox}
\usepackage{xcolor}
\usepackage{listings}
\usepackage{colortbl}
\usepackage{pbox}
\usepackage{graphicx}
\usepackage{arydshln}
\usepackage{bm}
\usepackage{amsmath}
\usepackage{makecell}
\usepackage{url}
\usepackage{hyperref}
\usepackage[normalem]{ulem}
\usepackage{float}
\usepackage{setspace}
\usepackage{enumitem}
\usepackage{balance}
\usepackage[font=small]{subcaption}
\usepackage{caption}

\usepackage[framemethod=tikz]{mdframed}

\AtBeginDocument{%
  }

\newcommand{\sysname}{\textsf{ByteHouse}\xspace}
\newcommand{\dc}{\textsf{CrossCache}\xspace}
\newcommand{\bkv}{\texttt{ByteKV}\xspace}
\newcommand{\fs}{\textsf{NexusFS}\xspace}
\newcommand{\ff}{\textsf{Sniffer}\xspace}

\newcommand{\hi}[1]{\vspace{.25em} \noindent {\bf #1} }
\newcommand{\bfit}[1]{\textbf{\textit{#1}}}
\newcommand{\zxh}[1]{\textcolor{blue}{#1}}

\newcommand{\blue}[1]{\textcolor{black}{#1}}

\theoremstyle{plain}
\newtheorem*{example*}{Example}

\newcommand{\myskip}{\vspace{0.5em}}
\definecolor{mygrey}{RGB}{230,230,240}

\definecolor{commentgreen}{RGB}{0,128,0}
\definecolor{keywordblue}{RGB}{0,0,255}
\definecolor{stringred}{RGB}{255,0,0}
\definecolor{backgroundgray}{RGB}{240,240,240}
\definecolor{numberorange}{RGB}{255,165,0}
\definecolor{preprocessorpurple}{RGB}{153,50,204}
\definecolor{mykeywords}{RGB}{153, 0, 0}

\copyrightyear{2026} 
\acmYear{2026} 
\setcopyright{cc} 
\setcctype{by} 
\acmConference[SIGMOD Companion '26]{Companion of the International Conference on Management of Data}{May 31-June 05, 2026}{Bengaluru, India} 
\acmBooktitle{Companion of the International Conference on Management of Data (SIGMOD Companion '26), May 31-June 05, 2026, Bengaluru, India} 
\acmDOI{10.1145/3788853.3803080} 
\acmISBN{979-8-4007-2450-3/2026/05}

\begin{document}

\title{\sysname:~ByteDance's Cloud-Native Data Warehouse for Real-Time Multimodal Data Analytics}

\author{Yuxing Han}
\affiliation{%
 \institution{ByteDance}
 \country{Shanghai, China}
 }
\email{hanyuxing@bytedance.com}

\author{Yu Lin}
\affiliation{%
 \institution{ByteDance}
 \country{Hangzhou, China}
 }
\email{linyu.michael@tiktok.com}

\author{Yifeng Dong}
\affiliation{%
 \institution{ByteDance}
 \country{Beijing, China}
 }
\email{dongyifeng@bytedance.com}

\author{Xuanhe Zhou}
\affiliation{%
 \institution{Shanghai Jiao Tong Univ.}
 \country{Shanghai, China}
 }
\email{zhouxuanhe@sjtu.edu.cn}

\author{XinDong Peng}
\affiliation{%
 \institution{ByteDance}
 \country{Beijing, China}
 }
\email{pengxindong@bytedance.com}

\author{Xinhui Tian}
\affiliation{%
 \institution{ByteDance}
 \country{Beijing, China}
 }
\email{tianxinhui@tiktok.com}

\author{Zhiyuan You}
\affiliation{%
 \institution{ByteDance}
 \country{Hangzhou, China}
 }
\email{youzhiyuan@bytedance.com}

\author{Yingzhong Guo}
\affiliation{%
 \institution{ByteDance}
 \country{Beijing, China}
 }
\email{guoyingzhong.z@bytedance.com}

\author{Xi Chen}
\affiliation{%
 \institution{ByteDance}
 \country{Shanghai, China}
 }
\email{max.chenxi@bytedance.com}

\author{Weiping Qu}
\affiliation{%
 \institution{ByteDance}
 \country{Shanghai, China}
 }
\email{quweiping@bytedance.com}

\author{Tao Meng}
\affiliation{%
 \institution{ByteDance}
 \country{Hangzhou, China}
 }
\email{jingpeng.mt@bytedance.com}

\author{Dayue Gao}
\affiliation{%
 \institution{ByteDance}
 \country{Beijing, China}
 }
\email{gaodayue@bytedance.com}

\author{Haoyu Wang}
\affiliation{%
 \institution{ByteDance}
 \country{Shanghai, China}
 }
\email{wanghaoyu.0428@tiktok.com}

\author{Liuxi Wei}
\affiliation{%
 \institution{ByteDance}
 \country{Singapore, Singapore}
 }
\email{weiliuxi@bytedance.com}

\author{Huanchen Zhang}
\affiliation{%
 \institution{Tsinghua University}
 \country{Beijing, China}
 }
\email{huanchen@tsinghua.edu.cn}

\author{Fan Wu}
\affiliation{%
 \institution{Shanghai Jiao Tong Univ.}
 \country{Shanghai, China}
 }
\email{wu-fan@sjtu.edu.cn}

\begin{abstract}

\begin{sloppypar}

With the rapid rise of intelligent data services,
modern enterprises increasingly require efficient, multimodal, and cost-effective data analytics infrastructures. However, in ByteDance's production environments, existing systems 
fall short due to limitations such as I/O-inefficient multimodal storage, inflexible query optimization (e.g., failing to optimize multimodal access patterns), and performance degradation caused by resource disaggregation (e.g., loss of data locality in remote storage). To address these challenges, we introduce \sysname~(\textit{\blue{\url{https://bytehouse.cloud}}}), a cloud-native data warehouse designed for real-time multimodal data analytics. The storage layer integrates a unified table engine that provides a two-tier logical abstraction and
physically consistent layout, SSD-backed cluster-scale cache (\dc) that supports shared caching across compute nodes, and virtual file system (\fs) that enable efficient local access on compute nodes. The  compute layer supports analytical, batch, and incremental execution modes, with tailored optimizations for hybrid queries (e.g., runtime filtering over tiered vector indexes). The control layer coordinates global metadata and transactions, and features an effective optimizer enhanced by historical execution traces and AI-assisted plan selection. Evaluations on internal and standard workloads show that \sysname achieves significant efficiency improvement over existing systems.
\end{sloppypar}
\end{abstract}

\renewcommand{\shortauthors}{Yuxing Han et al.}




\maketitle

\section{Introduction}
\label{sec:intro}

\begin{figure}[!t]
  \centering
  \includegraphics[width=\linewidth]{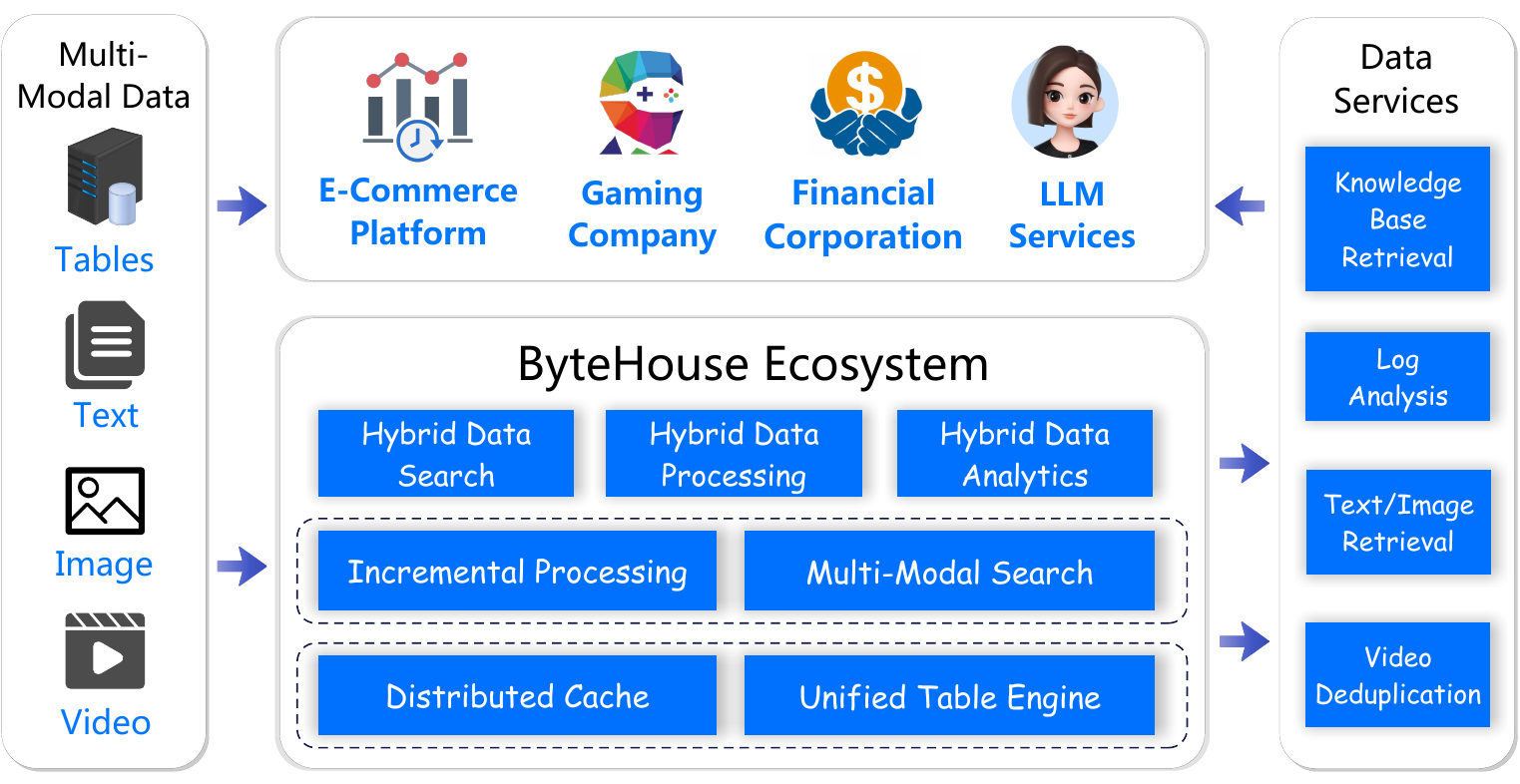}
%  \vspace{-1.75em}
  \caption{{\sysname provides real-time multimodal data analytics for over 400 services across ByteDance Cloud.}}
  \label{fig:intro}
\end{figure}

Real-time multimodal data analytics
is becoming increasingly important for modern enterprises (see Figure~\ref{fig:intro}).
Across ByteDance Cloud, a diverse set of applications, including e-commerce platforms, gaming services, financial systems, and AI-powered applications, operates over large and heterogeneous data modalities. These applications rely on analytical services such as operational dashboards, real-time log analysis, and semantic retrieval over text and image data to obtain timely and reliable insights. For instance, the code assistant Trae~\cite{gao2025trae} requires millisecond-level semantic retrieval across source code, documentation, and execution traces.

\begin{figure*}[!t]
  \centering
  \includegraphics[width=.95\linewidth]{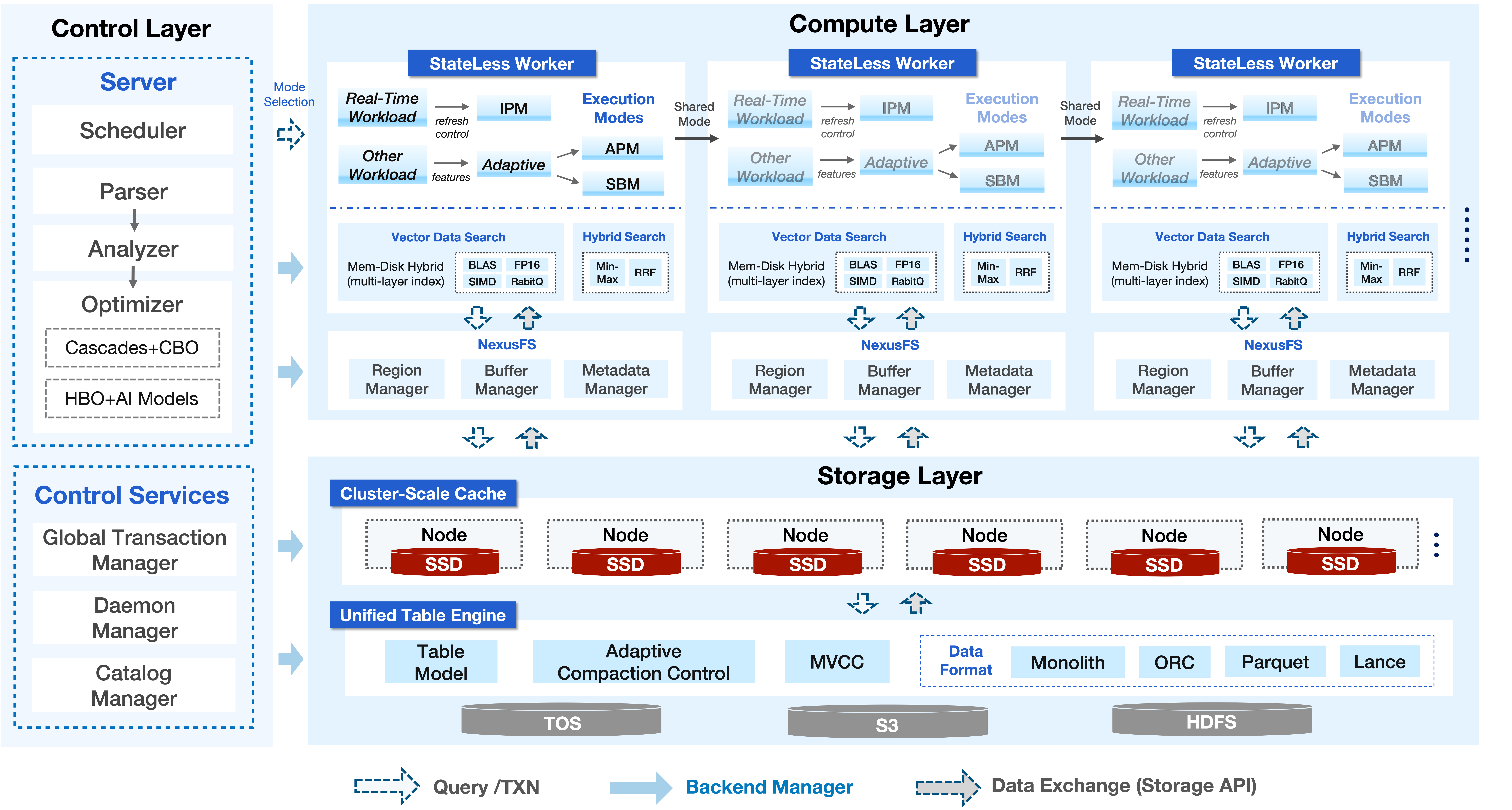}
  \vspace{-1.5em}
  \caption{\sysname Architecture.}
  \vspace{-1.5em}
  \label{fig:arch}
\end{figure*}

\begin{sloppypar}
Existing systems (e.g., data warehouses~\cite{redshift,PolardbMP,GaussDB,AnalyticDB,hologres,doris,starrocks,snowflake}, vector databases~\cite{wang2021milvus,pgvector}) have limitations in meeting these demands. First, many analytical engines~\cite{AnalyticDB,hologres} offer limited computation optimization for such multimodal workloads. For instance, they lack fine-grained incremental processing at operator level (e.g., delta-aware computation) and the data level (e.g., handling newly ingested rows). Additionally, their query optimizers are typically based on static rule- or cost-driven strategies and lack adaptive mechanisms that optimize evolving query patterns (e.g., text–vector fusion, scalar-constrained vector search) based on historical executions. Existing vector databases~\cite{wang2021milvus} and plugin-based solutions~\cite{pgvector} also struggle to efficiently optimize such hybrid queries. 
%First, many systems~\cite{AnalyticDB,hologres} provide limited support for realizing real-time multimodal workloads: (1) cannot support incremental processing at both the operator (e.g., delta-aware computation) and data level (e.g., newly ingested rows); and (2) rely on static, rule- or cost-based optimization and lack adaptive mechanisms to learn from historical executions and generalize across evolving query patterns, such as hybrid text–vector joins or multimodal data fusion. Existing vector databases~\cite{wang2021milvus} and plugin-based solutions~\cite{pgvector} also fall short in optimizing such hybrid queries. 
Second, many cloud-native architectures remain bottlenecked by remote object storage. Although disaggregated storage enables elasticity, it introduces significant latency for sparse or point-access workloads. Even with enhancements like vectorized execution or RDMA-backed memory sharing~\cite{PolardbMP,GaussDB}, performance degrades substantially for small or random reads, making real-time responsiveness difficult to sustain.
%Furthermore, the storage engines of these systems often decouple file-level indexes from data and scatter files across object-store shards~\cite{schulze2024clickhouse,snowflake}, which leads to substantial I/O amplification and additional consistency maintenance.
\blue{Furthermore, in many cloud-native analytical systems, file-level metadata and auxiliary indexing structures are maintained separately from base data files and stored in distinct object-store locations~\cite{schulze2024clickhouse,snowflake}.
Because data files are distributed across object-store partitions and accessed via remote metadata lookups, query processing may incur additional I/O amplification and metadata consistency overhead.
%, particularly under selective workloads.
}
%This fragmentation not only introduces redundant seeks and metadata lookups but also increases the cost of maintaining index–data coherence, ultimately degrading the performance of hybrid workloads.

\end{sloppypar}

To address these challenges, we propose \sysname, a cloud-native analytical data warehouse built upon a shared-storage architecture, which has been refined through {\it hundreds of application scenarios and tens of thousands of users.}   %Today, \sysname operates at massive production scale, spanning tens of thousands of nodes, managing near-exabyte data volumes.
(1) The storage layer provides a unified table engine with a self-describing columnar file format (\ff) that natively eliminates external metadata/index dependencies. And \sysname provides an SSD-backed distributed cache plane (\dc) that can independently scale and close the performance gap of traditional compute–storage separation. Additionally, \sysname supports efficient local data access through an alignment-aware filesystem abstraction on each compute node. 
(2) The compute layer offers three execution modes (i.e., analytic pipeline, staged batch, incremental processing) to handle diverse workloads such as real-time analytical processing.
Moreover, we introduce hybrid retrieval operators (e.g., \texttt{RANK\_FUSION}) for weighted semantic–lexical scoring, scalar-predicate-aware optimizations (e.g., runtime filtering across joined tables), and a tiered vector index design tailored to different service requirements. (3) The control layer manages global metadata, transactions, placement, and scheduling across the distributed architecture. Additionally, the query optimizer supports history-based optimization (HBO), which reuses runtime statistics from past executions to refine selectivity, cardinality, and operator-cost estimates. Building on HBO, \sysname incorporates AI models to learn correlations among query structures, data distributions, and runtime behavior to generalize beyond previously observed plan fragments, such as for predicate pushdown selection and join side selection.

\hi{Contributions.}In summary, we make the following contributions:
\begin{sloppypar}
\noindent(1) We propose \sysname, a cloud-native, shared-storage data warehouse for real-time multimodal data analytics, supporting high-throughput ingestion and ultra-low latency hybrid queries.

\noindent(2) We design a vertically integrated storage layer with a unified table engine, self-describing file format, SSD-based chunk-level caching, and a buffer-managed virtual filesystem that reduces I/O overhead across both the storage and compute nodes.

\noindent(3) We propose a unified execution framework combining analytic, batch, and incremental modes, with fusion-based retrieval operators and hybrid query optimization for complicated workloads.

\noindent(4) Extensive evaluations on both public benchmarks and ByteDance's internal workloads show that \sysname outperforms existing systems on both analytical queries (over 25\% lower latency in ClickBench) and multimodal data processing (over 50\% higher throughput in Cohere and C4). 
%{\it Today, \sysname operates at massive production scale, with over 25,000 nodes deployed across clusters (the largest one with 2,400 nodes), managing EB-level data in total.}

\end{sloppypar}

\section{\sysname Architecture}

Figure~\ref{fig:arch} illustrates the architecture of \sysname, a cloud-native analytical warehouse built upon a shared-storage design that fully decouples the control, compute, and storage services.

\begin{sloppypar}
\hi{Control Layer.}This layer provides global coordination within the distributed framework. The Server handles SQL parsing, semantic analysis, hybrid plan optimization (by history execution statistics and regression models), and distributed scheduling, translating user requests into executable workflows. The Control Services govern global metadata, placement, and system-wide transaction coordination. The Catalog Manager stores versioned metadata in \bkv, exposing snapshot-consistent schemas, partitions, and index definitions across concurrent operations. The Global Transaction Manager issues globally ordered commit timestamps to ensure serializable transactions and consistent snapshot reads. The Daemon Manager orchestrates background maintenance tasks such as compaction and merge scheduling. 
\end{sloppypar}

\hi{Compute Layer.}The compute layer hosts a pool of stateless workers operating in a hybrid execution model with three modes: (1) \textit{Analytic Pipeline Mode (APM)} for distributed multi-stage execution with shuffle, gather, and broadcast exchanges; (2) \textit{Staged Batch Mode (SBM)} for long-running ETL with staged retries and shuffle persistence; (3) \textit{Incremental Processing Mode (IPM)} for delta-aware computation with lineage tracking and versioned operator maintenance.
All modes share a unified optimizer and runtime to enable seamless transitions. To support multimodal retrieval and analytics, the compute layer provides a spectrum of index structures, including Min-Max, Set, Bloom filters, and vector indexes~\cite{malkov2018hnsw,jayaram2019diskann}. \sysname additionally introduces hybrid retrieval operators (e.g., \texttt{RANK\_FUSION}), scalar-predicate-aware optimizations (e.g., runtime filters pushed into vector scans), and a tiered vector indexing design tuned for online, near real-time, and cost-sensitive services.

\hi{Storage Layer.}This layer provides persistent and scalable multimodal data management through three tightly integrated components. (1) The \textit{unified table engine} offers a two-tier logical abstraction (documents–chunks) and a physically consistent layout composed of stable and delta segments. It manages ingestion through a staging–flush write path,
% backed by \bkv, 
 maintains snapshot-consistent visibility via MVCC, and supports adaptive compaction and tiered point-lookup resolution essential for analytical workloads. 
 (2) The \textit{distributed cache plane} comprises SSD-backed cache nodes that partition data into fine-grained chunks, apply consistent hashing for balanced placement, and employ prefetching and asynchronous flushing to mitigate remote-storage latency and reduce I/O amplification. (3) The \textit{virtual filesystem abstraction} (\fs) unifies access to heterogeneous storage backends, such as TOS (internal object storage), HDFS, and local SSDs, under a single logical namespace. \fs provides alignment-aware region management, coordinated metadata lookup, and buffer orchestration to sustain high-throughput read/write paths from compute nodes. Persistent data is stored in columnar formats such as Parquet, ORC, Lance, and the self-describing \ff format, enabling efficient hybrid processing across structured, textual, and vector modalities. All compute–storage data exchange is performed through Arrow-based~\cite{apachearrow} interfaces for zero-copy transfer and high-throughput communication across the disaggregated architecture.

\iffalse
\subsection{Data Model in \sysname}

On top of this tabular foundation, \sysname introduces a two-tier logical abstraction for multimodal data organization.
Each multimodal dataset is represented as a single table comprising a collection of \textit{documents}, where each document denotes a logical entity such as an article, image, or video.
Each document is further decomposed into semantically coherent \textit{chunks},
which constitute the fundamental units for embedding-based retrieval and analytical operations.
Every record in the table is uniquely identified by a composite key (\textsf{document\_id}, \textsf{chunk\_id}).
This document-chunk abstraction builds on the partition-part organization, where the physical organization ensures efficient data management and the logical abstraction captures semantic structure across modalities.

\subsection{System Architecture}

\fi

%\vspace{-4pt}
\section{\sysname Storage}
\label{sec:storage} %The storage layer of \sysname establishes the foundation for efficient, consistent, and scalable data management across analytical and multimodal workloads.

This section presents the storage design of \sysname for managing large-scale multimodal data in the cloud. We first introduce the Unified Table Engine, which provides a two-tier logical abstraction and physically consistent layout. We then detail a self-describing file format that co-locates raw data, indexes, and metadata to accelerate operations like point lookups. Next, we describe the distributed cache plane (\dc), which mitigates I/O bottlenecks in disaggregated architectures. Finally, we discuss \fs, a virtual file system that unifies local disk caching with remote \dc access for efficient compute-side reads.

\subsection{Unified Table Engine}

\sysname increasingly serves heterogeneous analytical workloads that span traditional OLAP queries, real-time incremental refresh, and multimodal data analysis. From the perspective of the storage engine, OLAP queries rely on large-scale scans and aggregations over immutable columnar data; incremental pipelines require efficient ingestion and low-latency access to recently updated results;
and multimodal workloads often perform feature-level retrieval, such as fetching semantic vectors by primary or sort keys, where point-lookups and fine-grained access latency become critical.

Existing systems typically rely on separate engines to handle these diverse patterns, leading to duplicated metadata and inconsistent data freshness~\cite{gruenheid2025autocomp}. Instead, \sysname introduces a Unified Table Engine that combines these capabilities under a single abstraction. It provides a consistent table model, unified metadata and versioning, and transactional visibility across analytical, incremental, and multimodal workloads.

\subsubsection{Table Model Design} \label{sec:table_model}
\sysname introduces a table model that defines how multimodal data are logically represented and physically organized.  % under a single abstraction
Each table encapsulates a unified schema that can contain structured attributes, nested fields, and vector columns derived from unstructured modalities. 
All columns are versioned and tracked by logical descriptors that record data type, encoding, and statistical summaries. 
Tables define \emph{primary keys and sort keys}, which together enable efficient range pruning for analytical scans and provide low-latency point lookups for feature retrieval tasks common in multimodal workloads.

\hi{(1) Logical Table Design.} On top of this table foundation, we introduce a two-tier logical abstraction for multimodal data organization. 
Each multimodal dataset is represented as a single table comprising a collection of \emph{documents}, where each document denotes a logical entity such as an article, image, or video. 
Each document is further decomposed into semantically coherent \emph{chunks}, 
which constitute the fundamental units for embedding-based retrieval and analytical operations.
Every record in the table is uniquely identified by a composite key (\textsf{document\_id}, \textsf{chunk\_id}).

\hi{(2) Physical Table Design.} Each table is physically composed of \emph{stable segments} and \emph{delta segments} that collectively maintain a consistent version chain. 
Stable segments store immutable, columnar data optimized for analytical throughput, while delta segments capture recent physical updates, inserts, and compact feature refreshes. 
Multi-version concurrency control (MVCC) governs the visibility of data across stable and delta segments, ensuring that analytical queries always read from a consistent snapshot even as refresh operations update the underlying data.
Background merge and compaction tasks asynchronously incorporate delta segments into stable segments, maintaining high ingestion performance while progressively organizing historical data.

\subsubsection{Adaptive Compaction Control}
Furthermore, efficient compaction of these delta segments is essential for controlling storage costs and sustaining query performance in large-scale data warehouse storage systems~\cite{gruenheid2025autocomp,agiwal2021napa}. However, the challenge is that overly frequent compaction amplifies write I/O and inflates storage costs, especially in cloud-native deployments where each write incurs overhead from remote storage, and insufficient compaction leads to segment fragmentation and degraded scan locality over time instead. %, and increasing maintenance overhead
Thus, \sysname introduces an adaptive compaction control mechanism that modulates compaction intensity to resolve the (observed) accumulated delta segments. % Rather than relying on static thresholds, the mechanism continuously adjusts compaction aggressiveness according to the observed degree of fragmentation and recent write amplification, achieving a balance between ingestion throughput and I/O efficiency.

Specifically, let $N_\Delta$ denote the number of active delta segments, and $N^*$ the equilibrium level that characterizes a balanced compaction state.
\sysname regulates compaction aggressiveness by a \textbf{bounded linear controller}, 
where the compaction intensity $\alpha \in [0, 1]$ is defined as:

\vspace{-1em} 
\begin{equation}
    \alpha = \min\left(1, \max\left(0, k \cdot \left(\frac{N_{\Delta}}{N_*} - 1\right)\right)\right),
\end{equation}
%\vspace{-3em}
with $k$ controlling the sensitivity to segment accumulation.
The coefficient $\alpha$ governs the triggering frequency, merge batch size, and scheduling priority of background compaction tasks.
When the number of delta segments is near the equilibrium level, 
the controller keeps $\alpha$ low, resulting in conservative compaction that avoids redundant compaction work. 
As $N_\Delta$ increases, $\alpha$ increases linearly, progressively relaxing merge thresholds and enlarging compaction batches until the controller saturates at full compaction intensity.
This monotonic linear control provides smooth transitions between idle and aggressive compaction phases, preventing oscillation and maintaining steady ingestion performance.

\subsubsection{\blue{Tiered Storage Pipeline}}

To efficiently handle small and frequent row-level ingestion without sacrificing columnar efficiency, \sysname employs a two-stage write pipeline that decouples transient ingestion from durable columnar persistence~\cite{antonopoulos2019socrates,LambFVTVDB12}.

\hi{(1) Staging in Key-Value Store.} Incoming updates are first written to a lightweight \emph{staging area}, supported by a distributed key-value store at ByteDance, \bkv.
Each write is assigned a transaction identifier and recorded in a Write-ahead Log to ensure durability and atomicity.
The staging area serves as a short-lived buffer that accumulates recent changes in row-oriented format.

\hi{(2) Flush to Columnar Storage.} Once the data size or retention time exceeds predefined thresholds, \sysname flushes the buffered data to columnar storage, reorganizing it into compressed column segments. During this flush process, schema evolution and version visibility are preserved, allowing newly persisted data to integrate seamlessly with existing analytical files.

\myskip

\blue{To support low-latency point access, \sysname resolves lookups across both staging and columnar layers under snapshot isolation. 
The engine first probes the transactionally consistent staging store for the most recent visible version of a tuple and falls back to columnar data parts when necessary, leveraging part-level pruning metadata
% (e.g., min/max statistics) 
 to minimize I/O. 
This tiered resolution follows the general separation of write- and read-optimized storage and ensures consistent visibility without requiring full-part scans.}

\subsection{Self-Describing File Format}

Next we introduce the file format in \sysname. In traditional column stores, column data and auxiliary indexes are maintained as separate structures. While this organization is effective for large sequential scans, it forces point and small-range queries to open additional files, issue seeks, and perform metadata lookups, leading to substantial I/O amplification and metadata overhead in lookup-intensive or hybrid analytical workloads.
Furthermore, the common practice of representing \texttt{Array} columns as flattened offset-value pairs lacks efficient support for modern multimodel data, such as high-dimensional embedding vectors.

To overcome these limitations, \sysname adopts a self-describing file format (\ff) that co-locates data blocks with their file-level indexes and schema metadata within a single cohesive layout.
This way, \ff eliminates external index and metadata files and enables self-contained data access with minimal I/O overhead.

\begin{figure}[t]
  \centering
  \includegraphics[width=\linewidth]{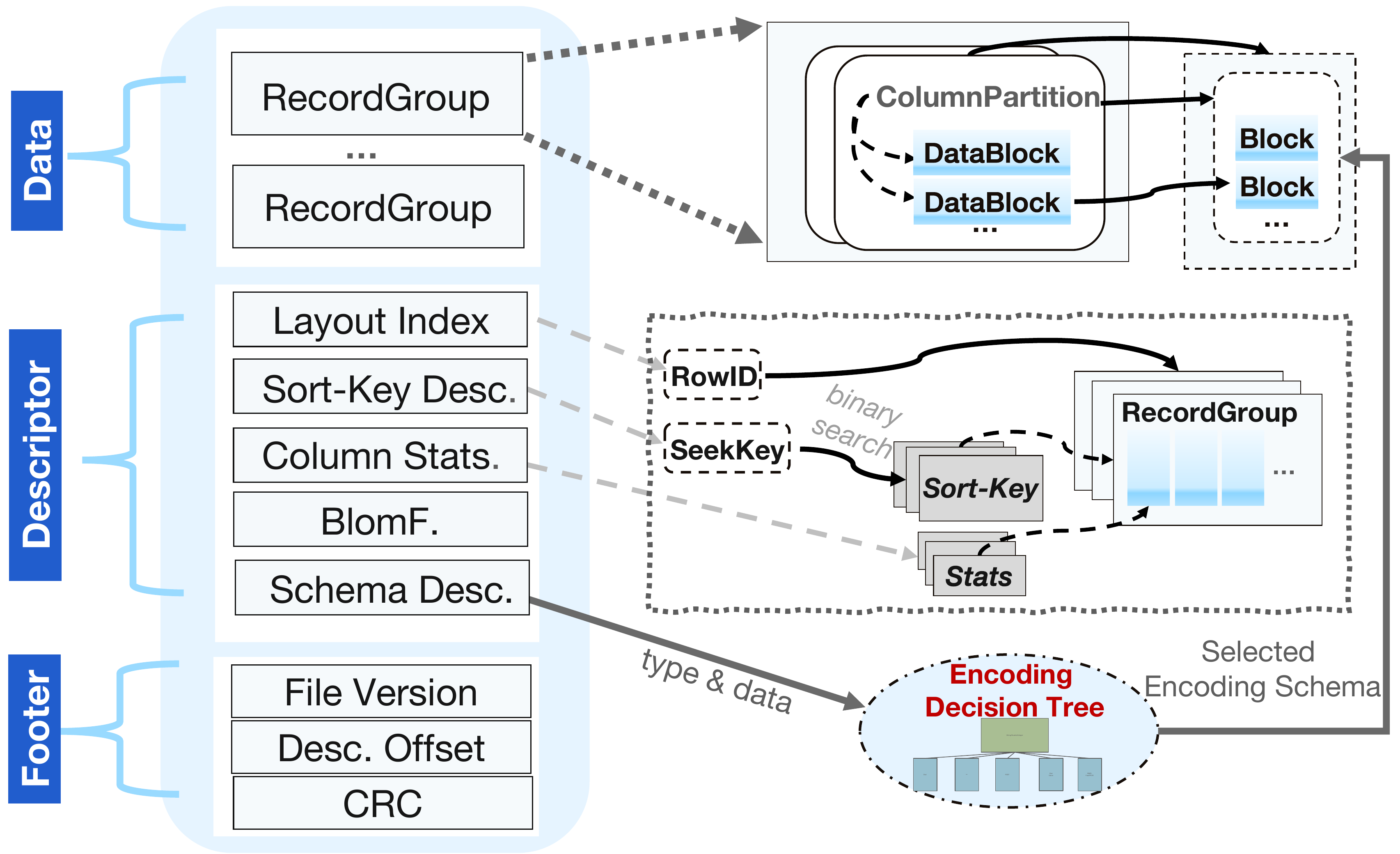}
  \vspace{-2em}
  \caption{\ff File Format.}
  \vspace{-.5em}
  \label{fig:sniffer}
\end{figure}

\vspace{-5pt}
\subsubsection{File Organization}

\ff is composed of Data, Descriptor, Footer regions in a hierarchical structure (see Figure~\ref{fig:sniffer}).

\noindent\textbf{(1) Data Region} stores compressed columnar data arranged in three levels: (i) A \emph{RecordGroup} corresponds to a batch of rows that share the same schema and partitioning key, analogous to a row group in Parquet or a stripe in ORC. (ii) Each record group contains multiple \emph{ColumnPartitions}, each representing a contiguous subset of a column's values within the group. (iii) Every column partition consists of one or more \emph{DataBlocks}, the smallest physical units that store compressed, type-specific blocks of values.

\noindent\textbf{(2) Descriptor Region} maintains all structural metadata required to interpret the Data Region efficiently: 
(i) The \textit{Layout Index} maps RecordGroups and ColumnPartitions to their physical DataBlocks and file offsets;
(ii)  The \textit{Sort-Key Descriptor} records ordering information for binary search and seek-key resolution; 
(iii)  The \textit{Column Statistics} store block-level summaries such as min/max and null counts for predicate pruning;
(iv)  The \textit{Bloom Filter} provides coarse-grained filtering for selective queries;
(v) The \textit{Schema Descriptor} defines the logical schema, data types, and encoding metadata (e.g., compression codec) used in the file.
\blue{Note, all metadata are file-scoped, which can be co-located with the storage layer within the file or maintained in a separate search acceleration layer~\cite{prammer2025towards}.}

\noindent\textbf{(3) Footer Region} serves as the global directory and integrity anchor of the file~\cite{zeng2026f3,vortex2024}.
It records the physical descriptor offsets, the file version, and the Cyclic Redundancy Check (CRC) values across the data and descriptor regions.
This design allows the reader to locate and validate all descriptors with a single footer access, enabling complete reconstruction of the file layout without relying on external catalogs.
Besides, the footer preserves version metadata to ensure backward compatibility and supports hierarchical integrity verification across regions.

Built upon this file format, we can accelerate fine-grained access to compressed columnar data by minimizing random I/O. At query time, the Layout Index, Sort-Key Descriptor, and Column Statistics collectively provide precise navigation over the hierarchical storage structure, allowing the engine to identify and load only the data blocks relevant to a given predicate, including key-based lookups.
For point lookups, the engine first identifies the target RecordGroup via the Sort-Key Descriptor, then consults the Layout Index to fetch the precise DataBlock offsets.  Because both descriptors are memory-resident,
each lookup typically requires only one metadata seek and one block read,
achieving \bfit{microsecond-level} latency even under high analytical concurrency.

\subsubsection{Column Encoding Schemes} Additionally, the column encoding methods are critical for the lookup efficiency (e.g., I/O, decoding overhead). Thus, \ff dynamically encodes each column according to its data type and statistical distribution~\cite{kuschewski2023btrblocks}.
During write time, \sysname samples each column and selects the encoding scheme that minimizes both storage footprint and decoding cost.
\ff integrates several state-of-the-art encoding schemes for different data characteristics:
Frame-of-Reference (FOR) combined with Bitpacking~\cite{lemire2015decoding} with narrow value ranges;
Run-Length Encoding (RLE) for low-cardinality columns with frequent repetitions;
Dictionary Encoding for categorical strings and enumerations, reducing both size and comparison overhead during predicate evaluation;
Finite-State Symbol Table (FSST)~\cite{boncz2020fsst} for general text or high-entropy string fields;
and Adaptive Losses Floating Point (ALP)~\cite{afroozeh2023alp} for floating-point columns, where precision can be adaptively reduced without compromising analytical accuracy.

\subsubsection{Format Optimization for Vector Data}
\begin{sloppypar}
In general column stores~\cite{schulze2024clickhouse,hologres,snowflake}, vectors are stored as flattened \texttt{Array} columns in the form of two companion buffers, i.e., an offset array indicating element boundaries and a value array storing all elements contiguously. While compact, this organization introduces several inefficiencies when applied to nested or vectorized data.
First, pooling all vector elements into a single value buffer prevents the system from maintaining per-vector statistics or applying compression schemes that operate at vector-block granularity.
Second, sparse or variable-length vectors must be represented using placeholder offsets or filler values, resulting in unnecessary storage space waste. % and reduced storage efficiency
\end{sloppypar}

To address this, \ff adopts the Length-and-Presence (L\&P) representation~\cite{melnik2020dremel} instead of the repetition/definition encoding used by Parquet~\cite{melnik2010dremel}.
L\&P attaches lightweight length and presence metadata to each vector, which enables \ff to isolate every embedding as an independent physical unit.
This structure enables the system to maintain vector-level statistics, such as value ranges, norms, and nullness, within the Descriptor Region.
Besides, it naturally captures sparse or variable-length embeddings without requiring padding or auxiliary offset structures.
Therefore, storage cost scales with actual content rather than with the declared dimensionality.
Since each vector is stored as a contiguous slice, block-oriented encodings such as FOR or ALP can be applied independently to each vector block.
This improves locality and enables SIMD-efficient decoding for dense numeric vectors.

\subsection{Cluster-Scale Caching on Storage Side}

\begin{sloppypar}

In cloud-native architectures, decoupling elastic compute from shared storage exposes a substantial performance gap when analytical workloads operate on cold or partially cached data.
As a result, query execution often becomes I/O-bound, with remote storage accesses introducing high latency.
These factors amplify tail stalls and frequently dominate end-to-end analytical response time.
Production traces from \sysname deployments show that small-block I/O can vary by more than an order of magnitude across percentiles, with over 70\% of execution time spent on cold reads in I/O-intensive workloads.
As the cluster scales, read amplification and tail latency increasingly constrain analytical performance.
\end{sloppypar}

%The root causes of remote I/O performance degradation vary across storage backends. In HDFS deployments, latency is inflated by NameNode lookups and block-location resolution; replication improves availability but does not materially reduce cold-read latency because each read still targets a single replica. On the other end, object storage (e.g., AWS S3) exposes high first-byte latency and limited per-request throughput~\cite{durner2023anyblob}, requiring hundreds of concurrent requests to saturate bandwidth and imposing nontrivial CPU and scheduling overhead.

\hi{Distributed Cache Design.} We design \dc, a distributed caching plane that mitigates remote-I/O bottlenecks by serving hot data from nearby nodes. \dc is composed of two main parts: (1) Cache Coordinators (CCs) that manage the global namespace and metadata, and (2) Cache Nodes (CNs) that provide local SSD-based caching and interact directly with different storage backends. Each cached file is divided into fixed-size blocks (12 MB by default), with block metadata and placement information maintained by the CNs. CNs handle both read and write operations.
For read paths, when a file is first accessed, the CN checks whether its metadata exists in the cache namespace; 
if absent, it retrieves the metadata from the backend, partitions the file into blocks, and registers them for subsequent access.
%During normal execution, requested data is served directly from \dc's local caches when available or transparently fetched and cached from the backend otherwise. 
For write paths, CNs buffer new data locally and coordinate block uploads once the buffers are filled. 
Small blocks are merged into unified files for distributed file systems 
or concatenated into larger objects for object stores.
CNs periodically report block mappings and  metadata to the CCs, which consolidate these updates to maintain global consistency across the cluster.

\hi{Fine-grained Chunk Caching.} To further mitigate I/O amplification and improve cache effectiveness under highly fragmented analytical access patterns, \dc adopts a {fine-grained chunk caching} strategy.
Each cached data block is further divided into 4 MB chunks, which are indexed in memory at each CN.
Contiguous chunks are appended sequentially to the SSD-resident block file, whereas non-contiguous chunks are temporarily buffered until they can be coalesced into a continuous range.
To address the limited write concurrency of storage backends such as HDFS (where each file admits only a single writer), \dc employs a \bfit{parallel flushing} mechanism.
Multiple CNs upload their completed block files concurrently as temporary objects, fully utilizing available network and I/O bandwidth.
A lightweight \textsf{concat} operation later merges these objects into a single backend file. 
This design substantially improves write throughput and reduces persistence latency under data-intensive workloads.

\vspace{-5pt}
\subsection{File System Abstraction on Compute Side}
While \dc enables shared caching across compute nodes, each compute worker still requires an efficient local access path to interface with it. 
\blue{In this context, a data fragment refers to a logical sub-range of a file requested by the execution engine.}

\blue{In many existing local caching mechanisms, these fragments are handled independently at arbitrary byte offsets~\cite{berg2020cachelib}.
When such requests do not align with the underlying storage block boundaries (e.g., SSD page size or object-store block size), the system must perform unaligned I/O, meaning that the requested byte range spans multiple physical blocks or requires partial-block reads.
Frequent unaligned small I/Os degrade throughput, waste SSD and network bandwidth, and undermine the effectiveness of prefetching, especially in scan-heavy analytical workloads.
}

\fs bridges this gap by introducing an {alignment-aware file-system abstraction} that unifies local disk caching and remote \dc access. It enforces end-to-end alignment across the remote data layout, on-disk cache regions, and in-memory buffers, ensuring that data is stored and accessed in coalesced and sequential units across all layers. Internally, \fs consists of three tightly integrated components: a region manager, a buffer manager, and a metadata manager.
The \bfit{region manager} partitions local disk space into fixed-size regions, typically 1 MB each, which are further divided into data segments that serve as the basic units of caching and I/O scheduling.
It maintains a global index mapping logical file offsets to cached segment locations and reclaims space through a FIFO-based eviction policy~\cite{yang2023fifo}.
The \bfit{buffer manager} governs in-memory caching through a fixed-size pool of segment-aligned buffers, applying a second-chance replacement policy to balance reuse and concurrency.
When a segment is retained for query processing, it can be directly exposed to the execution pipeline, enabling \textit{zero-copy} reads.
The \bfit{metadata manager} maintains a hierarchical namespace for all files accessed through \fs, using a two-level hash hierarchy
to track cached segments efficiently~\cite{Zhang2024PhatKV}. 
This structure supports constant-time lookups while minimizing memory overhead by selectively serializing inactive entries.

\section{Execution Engine}
\label{sec:compute}

\sysname provides three execution modes: the Analytical Pipeline Mode (APM) for low-latency interactive analytics, the Staged Batch Mode (SBM) for high-throughput offline pipelines, and the Incremental Processing Mode (IPM) for continuously ingesting new data.
APM and SBM are coordinated by a mode-adaptive execution framework that uses an AI-assisted selector to extract runtime features, predict resource demand, and route each query to the more efficient mode.
For IPM, \sysname incorporates a rule-based refresh controller that adjusts the refresh interval using a lightweight stabilization policy to balance freshness and maintenance cost.

\subsection{Execution Modes} \label{sec:exec_mode}
This subsection presents the three execution modes of \sysname.

\subsubsection{Analytic Pipeline Mode}\label{sec:APM}

To accelerate analytical workloads, \sysname supports an Analytic Pipeline Mode (APM) that optimizes query planning and resource allocation for low-latency analytical processing.
At its core, APM adopts a vectorized, pipeline-parallel execution model that processes columnar data in CPU-efficient batches, achieving high instruction throughput with minimal scheduling overhead.
APM incorporates several operator-level optimizations to accelerate core analytical primitives.

For aggregations, it employs an \textit{adaptive} mechanism that samples early input to estimate grouping-key cardinality and the reduction ratio, dynamically choosing between partial aggregation or direct shuffling followed by merge aggregation to avoid unnecessary local grouping and redundant merge computation.
For joins, APM applies \textbf{runtime filters} generated by build-side operators and pushes them down to probe-side scans, enabling early elimination of non-matching join keys and significantly reducing both probe-side data volume and the irregular, non-sequential access patterns that typically arise during join probing.

To maintain stable pipeline execution under high concurrency, APM integrates a \textbf{credit-based flow control} scheme in which each downstream operator grants a bounded number of processing credits to its upstream producers.
This regulates data production and buffer occupancy, preventing excessive queuing and sustaining steady pipeline throughput.
APM also supports an \textbf{ordered consumption} mechanism that preserves the order of intermediate results when the upstream operator already produces ordered output.
The consumer incrementally merges incoming ordered segments as they arrive, avoiding a full materialization-and-sort step and reducing overall query latency.

\subsubsection{Staged Batch Mode}\label{sec:SBM}

While APM targets interactive and low-latency analytical workloads, many production scenarios, such as large-scale ETL jobs and LLM-oriented data normalization pipelines, require long-running and throughput-oriented execution.
To support these workloads, \sysname provides a Staged Batch Mode (SBM), a stage-based execution framework built for stability, recoverability, and sustained throughput.
SBM divides a query plan into a sequence of stages separated by \textit{exchange} boundaries. 
Each stage is decomposed into parallel tasks that operate on disjoint data partitions, and a stage begins execution only after all of its upstream dependencies have completed.
Within a stage, tasks can \textbf{materialize intermediate results} to temporary storage, allowing downstream tasks to retrieve their inputs from local or remote spill files based on data locality.
These materialized outputs act as lightweight checkpoints and enable \textbf{task-level retries}, where failed tasks can be relaunched without restarting the entire stage, significantly reducing recovery time and operational overhead.
SBM further supports \textbf{elastic parallelism}: input partitions assigned to a compute worker can be processed in multiple batches, allowing the system to flexibly increase task parallelism while keeping each task's memory footprint small.

\subsubsection{Incremental Processing Mode}\label{sec:IPM}
To support near real-time analytical workloads, IPM maintains continuously updated query results as base tables evolve.
Unlike APM and SBM, which execute queries by recomputing their inputs in a full scan or a stage-based batch pipeline, IPM maintains operator state across refresh cycles and evaluates only the data deltas produced since the previous materialization point. 
This incremental model significantly reduces computation cost and end-to-end latency while preserving transactional consistency~\cite{akidau2023s,sotolongo2025streaming,budiu2022dbsp,wang2020tempura}.
Realizing IPM requires coordinated changes across the query processing stack.

\noindent \underline{\textbf{SQL Semantics and Triggering}}: IPM remains fully compatible with standard syntax and extends DML statements with \textit{refresh interval} annotations that specify the desired maintenance frequency.
Shorter intervals provide fresher results at higher compute cost, whereas longer intervals reduce system load while maintaining consistency guarantees.

\noindent \blue{\underline{\textbf{Row-Level Lineage and Delta Model}}: IPM introduces \textit{row-level lineage tracking} to support
deterministic and fine-grained incremental maintenance.
Unlike simple modification flags that merely indicate whether a row has changed, \sysname associates each tuple with two immutable metadata fields: i) \texttt{tuple\_key}, which uniquely identifies the logical identity of a row across versions; ii) \texttt{update\_seq}, a monotonically increasing sequence number that records the version order of updates for that logical row.}

\blue{During incremental execution, operators consume streams of delta tuples annotated with this lineage metadata.
The lineage information is systematically propagated along the operator pipeline to preserve traceability.
For example, in an inner join, the output tuple inherits a composite lineage derived from the participating base tuples.
 When a base tuple is updated or deleted, the engine uses its \texttt{tuple\_key} to locate and retract previously materialized join results associated with earlier \texttt{update\_seq} values. This ensures correctness under incremental recomputation without requiring full re-scans of base tables.}

\noindent \underline{\textbf{Operator-Level Delta Execution}}: 
%IPM provides \textit{incremental operators} that process delta inputs and update operator state without reprocessing full base tables.
\blue{All incremental operators consume and emit delta records of the form <\texttt{tuple\_key}, \texttt{update\_seq}, \texttt{op}>, where \texttt{op} $\in \{insert, delete\}$.
A logical update generates a delete delta for the previous version, followed by an insert delta for the new version. Upon receiving a delete delta, an operator locates previously materialized state entries (e.g., stored in state tables) derived from the given \texttt{tuple\_key}, removes their contribution, and propagates a corresponding delete delta downstream. 
This uniform delta protocol ensures compositional retraction across operators.}

\blue{For aggregation, IPM maintains a state table keyed by grouping attributes, where each entry materializes the current aggregate state, including partial aggregates and any derived results.
Incoming deltas trigger state lookups and either apply new values or retract obsolete ones, while groups whose aggregate values (e.g., \texttt{COUNT}) drop to zero are eliminated through lightweight deletions.
Aggregate maintenance is specialized according to algebraic properties.
For natively retractable aggregates, such as \texttt{COUNT}, \texttt{SUM}, and \texttt{AVG}, IPM adopts fully incremental implementations. 
Insert deltas are accumulated into the maintained state, whereas delete deltas are subtracted from their prior contributions.
In contrast, aggregates that do not efficiently support retraction under arbitrary update streams, 
such as \texttt{MIN}, \texttt{MAX}, are handled using a fallback strategy. 
IPM retains the original per-group values within the operator state to enable recomputation confined to the affected group.
When a delete delta invalidates the current aggregate result, for example, by removing the current minimum, the operator recomputes the aggregate over the affected group.
This design preserves correctness while confining recomputation to impacted groups, trading additional memory overhead for bounded recomputation cost.}

For inner joins, the optimizer rewrites each join into three delta subqueries that combine the left delta with the right base, the right delta with the left base, and both deltas together.
These subqueries execute in parallel over versioned inputs managed by the GTM to ensure snapshot consistency.
Their outputs are then unified through a lineage-based reconciliation step that merges overlapping tuples by \texttt{tuple\_key} and modification sequence.

 For left or right outer joins, IPM augments the plan with correction terms that maintain the correct semantics of null extension. 
The system tracks whether each row currently has at least one matching partner and emits updates whenever this match status changes.
When an unmatched row gains a match, its null extended output is withdrawn, and the matched rows are emitted; when the final match disappears, the matched rows are retracted, and a single null extended row is produced.

\subsection{Adaptive Execution and Refresh Control} \label{sec:mode_selection}

To balance responsiveness, throughput, and data freshness, \sysname employs a learning-based mode selector that adapts query execution to workload characteristics, and a rule-based refresh controller that schedules incremental maintenance for IPM.

\subsubsection{Adaptive Mode Selection Between APM/SBM}\label{sec: adaptive_mode_selection}
The learning-based mode selection framework performs predictive mode selection using a lightweight regression model.
Upon query submission, a feature extractor parses the SQL statement and derives a composite query feature vector from the query plan.
The vector consists of three parts: 1) query-level features, such as query length and number of views, 2) access-pattern features, including a one-hot encoding of referenced tables, and 3) plan-structural features obtained by bottom-up traversal of plan nodes, where each node is represented by an $M$-dimensional structural vector.
This feature vector is fed into a regression model trained from historical executions.
The model jointly predicts query latency, CPU usage, and memory demands, which are then mapped to mode classes (i.e., APM or SBM) using \textit{multi-dimensional percentile} thresholds derived from the cluster's recent workload statistics.
These thresholds are continuously recalibrated to capture runtime variations in resource contention and query concurrency, ensuring balanced mode selection between responsiveness and throughput.
The prediction model is periodically retrained with recent execution statistics to maintain accurate and adaptive resource estimation.

\subsubsection{Rule-based Refresh Control}\label{sec:refresh_control}
Currently, IPM is explicitly triggered by users when continuous or periodic data maintenance is required.
Automatically determining an appropriate refresh interval is crucial for incremental processing, as manually configuring a fixed refresh interval can lead to either excessive recomputation or stale data under varying workloads.
To address this, \sysname employs a window-based stabilization policy that smooths transient fluctuations in maintenance cost and ensures convergence toward a steady refresh frequency.
Instead of relying solely on the most recent observation, the policy maintains a sliding window of execution times from recent refresh rounds, denoted by $T_1, T_2, ..., T_N$.
It estimates the average incremental maintenance cost as:
\begin{equation}
    T_{avg} = \frac{1}{N} \sum^N_{i=1} T_i,
\end{equation}
where $N$ is typically small (e.g., 3-5) to balance stability and responsiveness.
The next refresh interval is computed as:
\begin{equation}
\Delta t = \min\left( \max\left( k \times T_{\text{last}}, \Delta t_{\min} \right), \Delta t_{\max}(U) \right),
\end{equation}
\noindent which explicitly restricts the interval to lie within a lower bound $\Delta t_{min}$ and an upper bound $\Delta t_{max}(U)$.
The scaling factor $k$ prevents overly frequent maintenance, while $\Delta t_{min}$ protects the system from thrashing.
The upper bound $\Delta t_{max}(U)$ adapts to real-time cluster utilization $U$: \sysname enlarges it under high CPU or I/O load to reduce contention, and contracts it under low load to improve freshness.
A simple linear model is used in practice:
\begin{equation}
    \Delta t_{max}(U) = \Delta t_{base} \times (1 + \alpha U),
\end{equation}
where $\Delta t_{base}$ denotes the nominal upper bound under idle conditions, and $\alpha$ controls the sensitivity to load variation.
By averaging recent refresh costs and adapting to real-time cluster utilization, this policy absorbs transient spikes, prevents runaway growth of $\Delta t$,  and converges to a stable refresh frequency that balances data freshness with overall system efficiency.

\section{Query Optimization} \label{sec:optimize}
This section describes \sysname's query optimization techniques.
Section~\ref{fig:opt_cascades} describes the optimizer architecture and its Cascades-based rewrite and enumeration framework, while Section~\ref{fig:opt_ai} introduces AI-driven methods for predicate pushdown selection and join side selection.

\subsection{Cascades Optimizer}\label{fig:opt_cascades}
\sysname's query optimizer builds upon the Cascades framework~\cite{graefe1995cascades,graefe1993volcano,soliman2014orca} and extends it with a unified cost-based model (CBO) that reasons about partitioning, sorting, and grouping properties during query optimization~\cite{zhou2010incorporating}.
Each operator is annotated with structural metadata describing data distribution and ordering, allowing the optimizer to derive and enforce these properties \textit{consistently} throughout plan enumeration.
To efficiently explore the search space of bushy join trees, 
\sysname adopts the branch partitioning top-down enumeration algorithm~\cite{fender2011femo}, which enumerates join combinations in constant time and eliminates the complex data structures required by earlier approaches.
The optimizer further incorporates a Magic Set rewriting~\cite{magicset,sereni2008adding} to enhance predicate selectivity propagation.
By replicating selective subplans into upstream operators, this transformation enables early filtering and significantly reduces intermediate results.
\sysname also employs a cost-based optimization for Common Table Expressions (CTEs), where the optimizer decides adaptively whether each CTE should be inlined, shared, or materialized based on its contextual reuse and query semantics~\cite{cte-optimization,silva2012exploiting}.

\subsection{\blue{Intelligent Optimization}} \label{fig:opt_ai}

To enhance the robustness and accuracy of query optimization in dynamic analytical workloads, \sysname adopts a history-based optimization (HBO) ~\cite{shankhdhar2024presto,galindo2003statistics}  that leverages runtime statistics collected from past query executions to guide future cost-based decisions. 
Each plan fragment is transformed into a canonical representation and assigned a hash value that serves as a lookup key in a lightweight history store. 
During query compilation, the optimizer performs hash-based matching to retrieve previously recorded statistics, such as predicate selectivities and join cardinalities, and incorporates them into the cost estimation.
While HBO improves the robustness and accuracy of cost estimation by reusing statistics from past executions, it remains fundamentally limited to plan fragments observed before and cannot fully accommodate new query patterns or evolving data distributions.

To overcome this limitation, \sysname incorporates regression models to learn correlations among query structures, data characteristics, and runtime behavior~\cite{neo,zhu2020flat,SunL19,saxena2023auto,ZhuCDCPWZ23,MarcusNMZAKPT19,trummer2019skinnerdb}. 
By capturing these learned relationships, \sysname can generalize beyond hash-based plan reuse and provide more accurate cost estimates for previously unseen workloads.
Beyond ByteCard~\cite{han2024bytecard}, which leverages probabilistic graph models~\cite{koller2009probabilistic} to refine cardinality estimation for base tables (e.g., filter selectivity and join size), we further \blue{deploy} learning-based optimizations targeting Predicate Pushdown Selection (PPS) and Join Side Selection (JSS).
%\blue{These techniques extend learning-driven estimation beyond base cardinalities to physical plan refinement decisions within the optimizer.}
\blue{Both optimizations have been validated in production deployments.}
%{The PPS technique has been rolled out across 35 clusters, covering 100 tables. On these deployed tables, the average I/O read volume has been reduced by more than 20\%.
%The JSS technique has been deployed on 3 clusters. In the corresponding workloads, join queries typically involve between 6 and 20 tables, and the resulting side selection accuracy exceeds 95\%.}

\begin{figure}[t]
\centering
\hspace{-2em}
\subfloat[\scriptsize Predicate Encoding]{
\label{fig:subfig:predicate}
\vspace{-2ex}
{\includegraphics[scale=0.11]{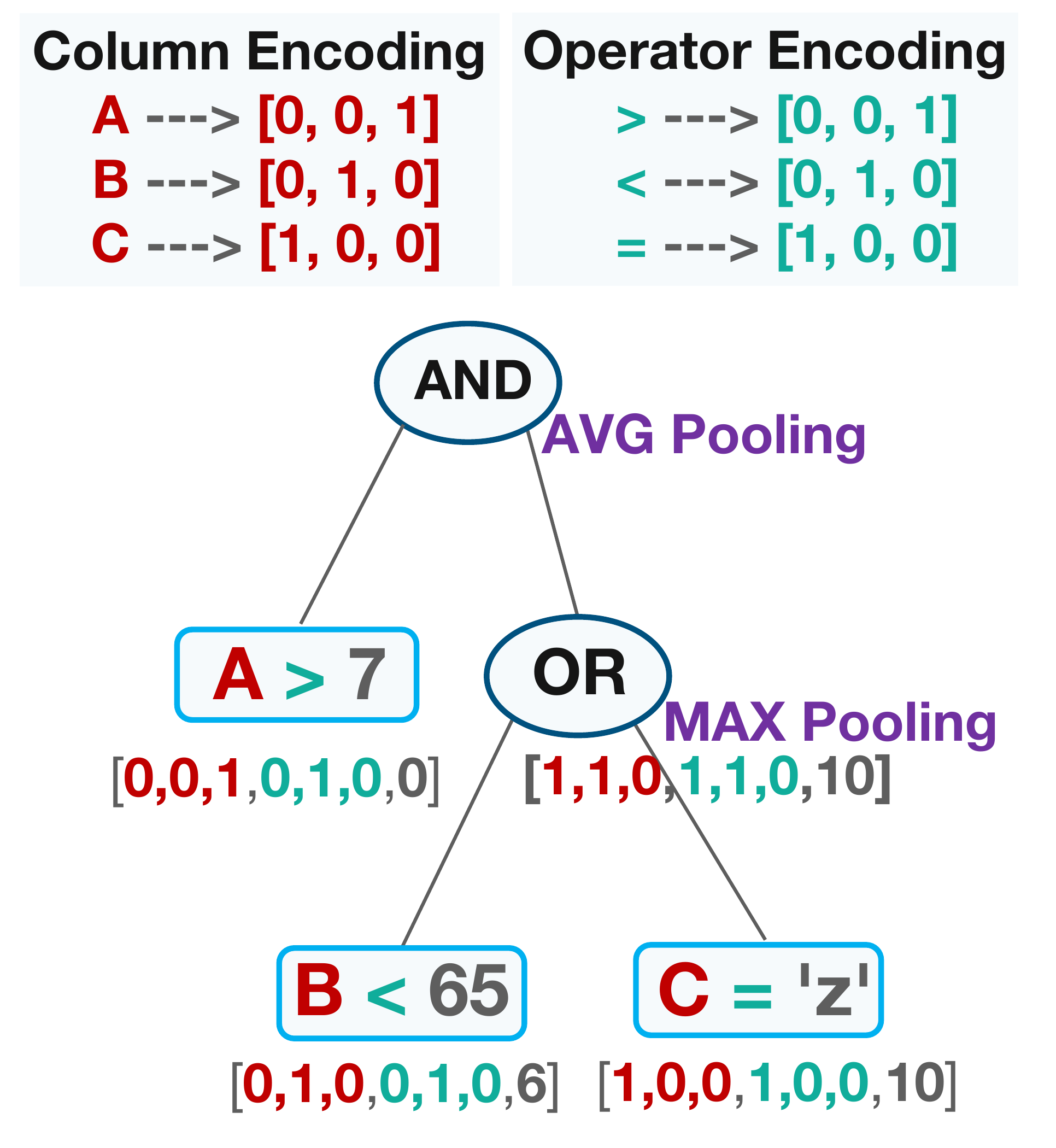}}
}
\hspace{-1em}
\subfloat[\scriptsize Join Side Modeling]{
\label{fig:subfig:join_model}
\vspace{-2ex}
{\includegraphics[scale=0.12]{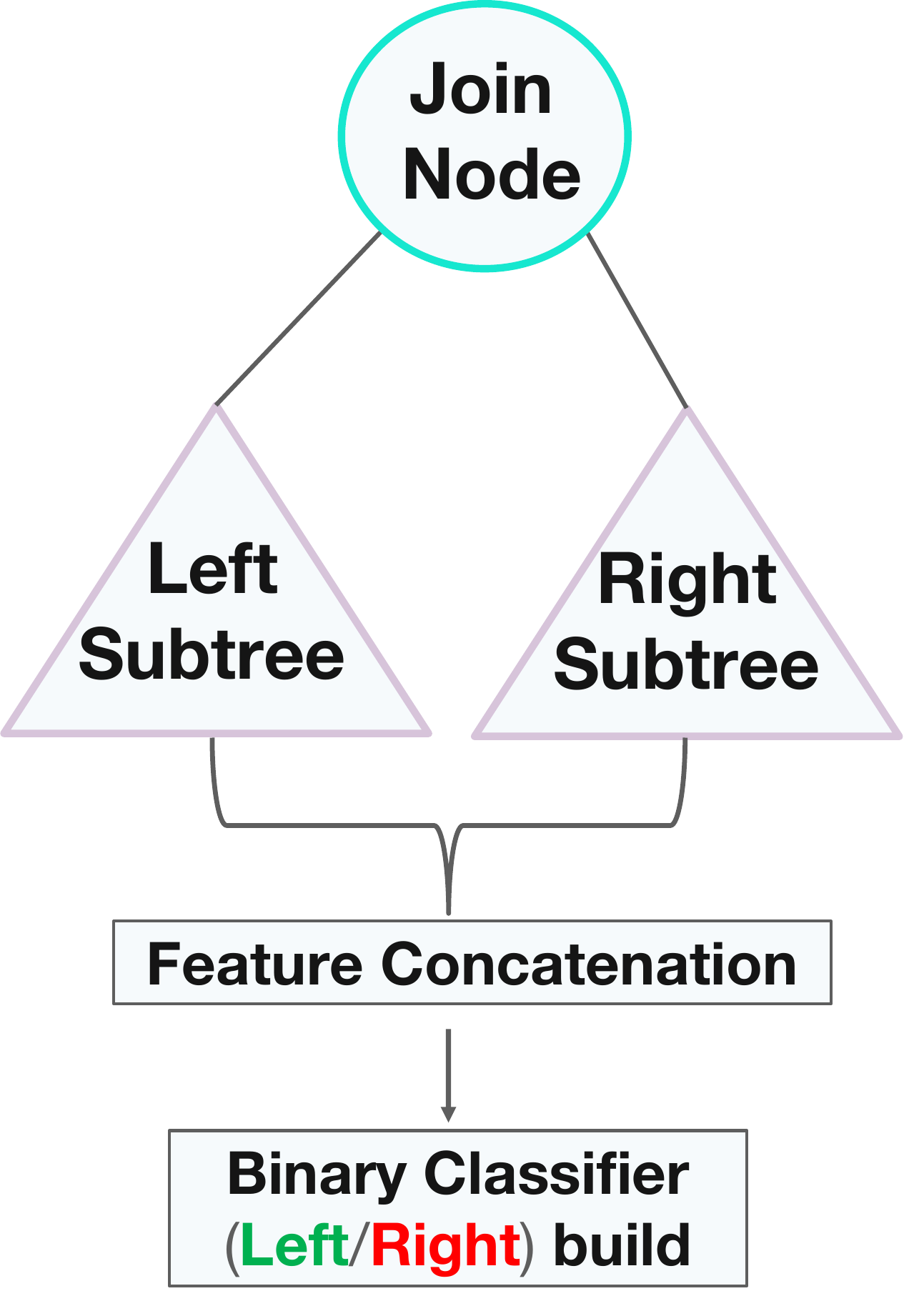}}
}
\hspace{-0.5em}
\subfloat[\scriptsize Bottom-Up Refine]{
\label{fig:subfig:join_reorder}
\vspace{-2ex}
{\includegraphics[scale=0.12]{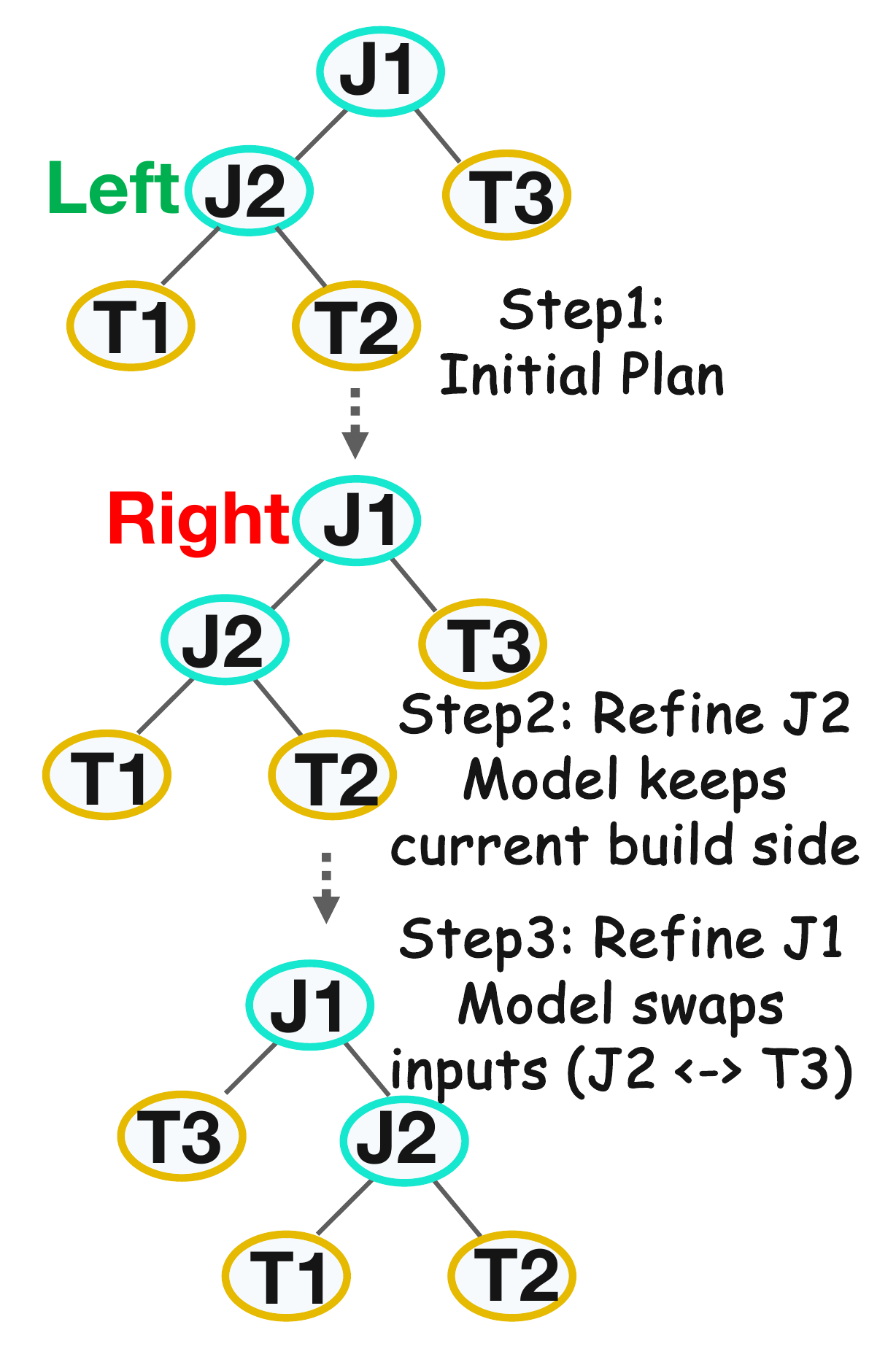}}
}

\vspace{-1em}
\caption{Intelligent Optimization by History Executions.}
\vspace{-.5em}
\label{fig:query_perf}
\end{figure}

\hi{{1) Predicate Pushdown Selection.}}
%Predicate pushdown~\cite{yan2023predicate,jeong2025upp} reduces query I/O by filtering data early, but determining which predicates to push down in complex analytical workloads is non-trivial.
\blue{While predicate pushdown~\cite{yan2023predicate,jeong2025upp} is commonly regarded as a straightforward optimization, its impact in \sysname is more nuanced due to disaggregated storage and multimodal execution. 
In particular, aggressively pushing down predicates whose evaluation cost is high relative to their selectivity benefit (e.g., vector similarity conditions or UDF-based filters) may increase remote I/O and processing overhead while precluding more effective pruning opportunities later in the plan. 
Therefore, predicate pushdown in \sysname requires cost-aware selection rather than indiscriminate application.}

We formulate this decision process as a supervised regression task for estimating predicate-specific I/O costs.
In the offline stage, we collect training samples of (\textsf{predicate}, \textsf{I/O cost}) pairs, and train a regression model that learns the mapping from predicate features to observed scan costs.
At runtime, \sysname leverages the trained model to evaluate candidate predicates and select the most cost-effective ones for pushdown.

To construct candidate predicates, \sysname decomposes each query's \textsf{WHERE} clause into top-level conjunctive components separated by \textsf{AND} operators, yielding independent candidates for pushdown evaluation.
Each candidate is represented as an Abstract Syntax Tree (AST) whose nodes correspond to comparison operators, column references, literal values, etc.
For node encoding, we apply one-hot encodings to categorical features such as comparison operators and column identifiers, while continuous features are discretized into value-domain buckets following prior work~\cite{MSCN,wu2020bayescard}.
For AST encoding,  conjunctions and disjunctions (\textsf{AND}, \textsf{OR}) are modeled as pooling functions~\cite{gholamalinezhad2020pooling} over their child encodings rather than as categorical tokens.
This approach can preserve the logical semantics of predicate composition.
Specifically, \textsf{OR} applies MAX pooling and \textsf{AND} applies AVG pooling.
The final predicate embedding is obtained by a postorder traversal across the AST that recursively aggregates node representations.
Figure~\ref{fig:subfig:predicate} presents an example of predicate encoding for: \textsf{(A > 7) AND (B < 65 OR C=`x')}.
Note, the underlying relational table contains three columns (\textsf{A}, \textsf{B}, \textsf{C}), and the analytical workload involves three comparison operators (\textsf{>}, \textsf{<}, \textsf{=});
columns \textsf{A} and \textsf{B} share a numeric domain of $[0, 100]$, and \textsf{C} is a categorical attribute with $10$ distinct values.

\hi{{2) Join Side Selection.}}
Join performance is highly sensitive to which input is chosen as the \textit{build} (hashed) side versus the probe side, as this decision governs hash-table size, memory pressure, and downstream I/O.
Traditional optimizers rely on cost models to select the build side, but their effectiveness hinges on accurate cardinality estimates and may degrade under skew or complex predicates.
In \sysname, we introduce a learning-based side-selection mechanism that refines the initial join physical plan produced by the optimizer.
As illustrated in Fig.~\ref{fig:subfig:join_model}, we formulate build/probe selection for a join node as a binary classification task that predicts which child should serve as the build input.
For each join node, the model constructs a feature vector by concatenating the learned representations of its left and right subtrees with features of the current join (e.g., join predicates, estimated selectivities, and row-width signals) in post-order;
this embedding is then fed to a classifier that outputs \textit{left-build} or \textit{right-build}.

During training, labels are derived offline by comparing the observed output cardinalities of the two subtrees: if the left subtree's cardinality is smaller, the label is \textit{left-build}; otherwise, \textit{right-build}.
To maintain semantic consistency across recursive subplans, we assume that when evaluating a join node, all its descendant joins have already been assigned correct side selections.
This assumption is enforced during training by restructuring query plans to satisfy it, and during inference by traversing the plan bottom-up (see Fig.~\ref{fig:subfig:join_reorder}), ensuring that lower-level joins are decided before higher-level ones.
In this manner, the model adaptively determines build/probe selection throughout the query plan, improving memory efficiency and hash-table locality while preserving overall query semantics.

\section{Multi-Modal Data Indexing and Hybrid Search}
\label{sec:retrieval}

In \sysname's production scenarios, there is a growing need to optimize the execution of multimodal search queries (e.g., text+vector retrieval in code assistant). While existing vector databases (e.g., Milvus-based solutions) are primarily designed for standalone similarity search, they lack tight integration with relational query execution. In contrast, \sysname builds upon a document-chunk abstraction (see Section~\ref{sec:table_model}) and extends standard SQL to support both vector and full-text search primitives within a single query. To efficiently execute such queries, \sysname introduces modality-aware optimizations, including runtime filtering, hybrid vector index designs, and multi-source result fusion strategies.

In the following, we introduce the design choices of vector indexes (in a hierarchical way) and hybrid data search in \sysname.

\hi{Multi-Layer Vector Indexes.} Within a single table, we determine the suitable index structures based on the service types. The candidate vectors undergo a multi-layer structure tailored for various latency, freshness, and cost trade-offs. First, a coarse index layer performs partition-level pruning using product quantization (PQ) and BLAS-accelerated centroid distance, avoiding unnecessary index traversal. The rest layers of the index are determined based on the service types: $(i)$ For latency-critical online services that demand high recall and millisecond-level responsiveness, \sysname adopts HNSW with scalar quantization (SQ), where vectors are pre-quantized into compact representations and organized in a navigable small-world graph. This allows bounded-depth traversal during query time, achieving fast search while maintaining recall. Under this setting, we can build the index asynchronously and decouple it from ingestion, ensuring minimal impact on write throughput.  $(ii)$ For near real-time services requiring high recall and rapid vector visibility (within seconds to sub-seconds), but with relaxed latency (e.g., 100ms-1s), \sysname adopts centroid-based partitioning during index construction, with each partition maintaining either full-precision vectors (IVFFlat), quantized vectors (IVFSQ), or PQ-based compression (IVFPQ). This enables fast ingestion-to-query cycles and moderate memory usage. $(iii)$ For cost-sensitive services, \sysname adopts DiskANN that stores full-precision graph structures on SSDs and caches routing metadata in memory. Additionally, we use optimized I/O prefetching for common queries and beam search to maintain latency under relaxed constraints. For even more relaxed workloads (e.g., over long-tail vectors order than six months), \sysname supports DiskIVFSQ to store quantized, partitioned vectors on disk, significantly reducing both memory and compute costs, making it suitable for archival data with minimal freshness and latency demands.

\begin{figure}[h]
  \centering
  \includegraphics[width=\linewidth]{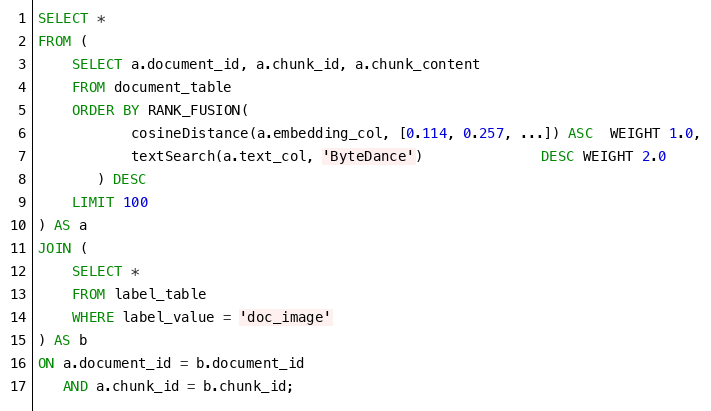}
  \vspace{-2em}
  \caption{Example Hybrid Data Search in \sysname.}
  \vspace{-.5em}
  \label{fig:hybrid_query}
\end{figure}

Upon carefully constructed vector indexes, \sysname further extends its support to more complicated hybrid data search queries that blend semantic, textual, and structured constraints. Consider the hybrid query shown in Figure~\ref{fig:hybrid_query}. The nested subquery retrieves the top-$K$ chunks based on a fused ranking over both semantic similarity (via vector-based cosine distance) and keyword relevance (via lexical text search). These two modalities are integrated through the \texttt{RANK\_FUSION} operator, which combines scores with configurable weights to balance semantic and lexical contributions. For the outer query, instead of issuing this vector-augmented query in isolation, we join its results with a filtered label table that selects only chunks annotated with a specific tag (e.g., \texttt{``doc\_image''}). This join is structured to enable scalar-predicate-aware optimizations. \sysname executes such hybrid queries in three main steps.

\begin{figure*}[t]
\centering
\subfloat[\small TPC-DS Benchmark (SF-1000).]{
\label{fig:subfig:tpcds}
{\includegraphics[scale=0.32]{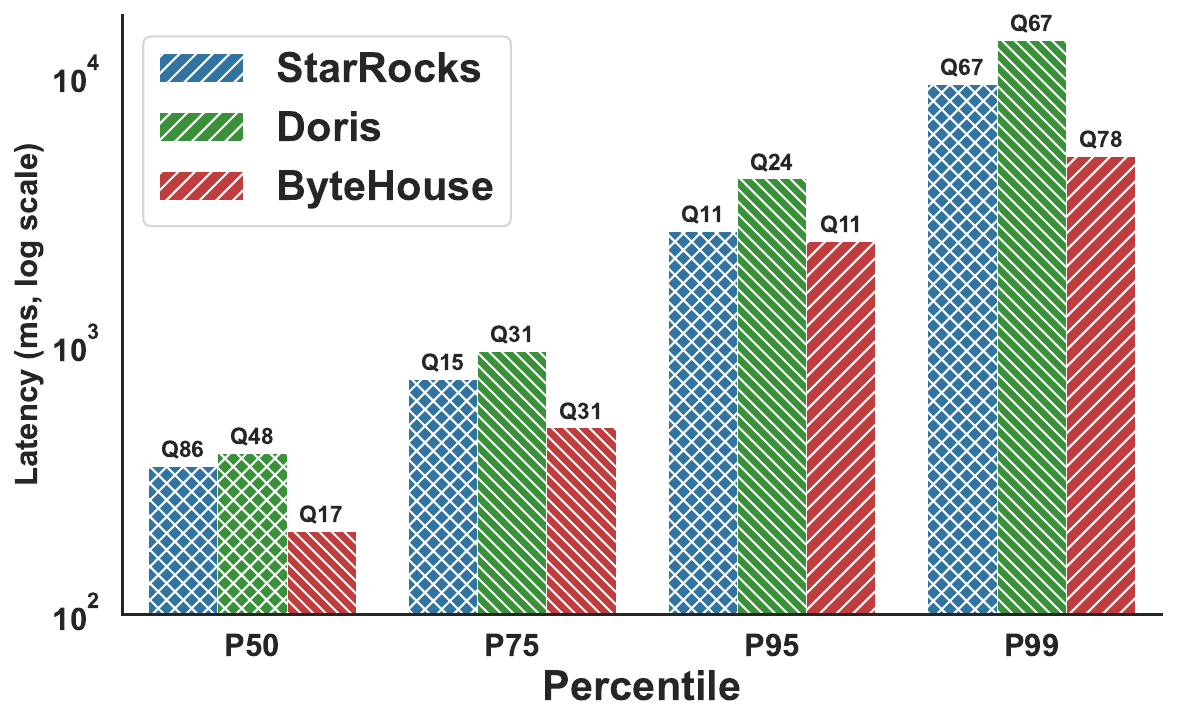}}
}
\subfloat[\small ClickBench Benchmark (Overall 25.4\% Latency Reduction).]{
\label{fig:subfig:clickbench}
{\includegraphics[scale=0.32]{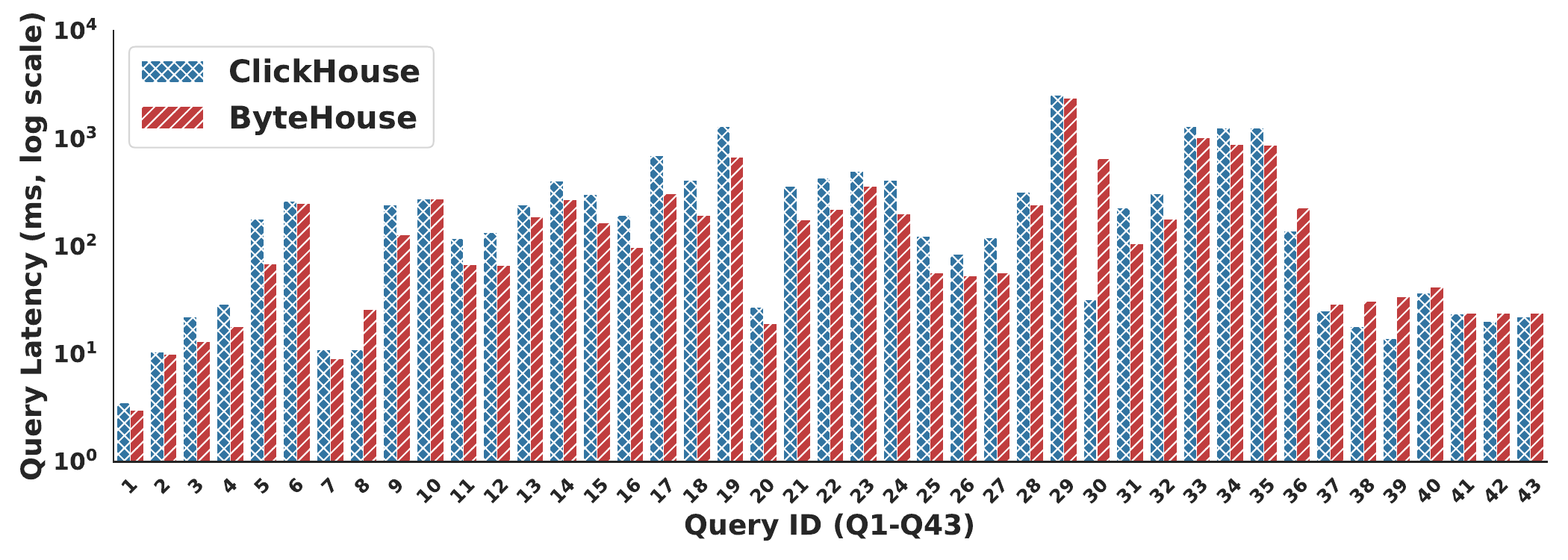}}
}
\vspace{-1em}
\caption{Comparative Results on Standard Benchmarks.}
\vspace{-.75em}
\label{fig:exp_standard_benchmark}
\end{figure*}

\hi{(1) Cross-Table Runtime Filtering.} First, \sysname conducts cross-table runtime filtering to prune the search space before initiating vector retrieval. The optimizer estimates the selectivity and optimal join order~\cite{niu2025blendhouse} (see Section~\ref{sec:optimize}). When the scalar-side table (e.g., label\_table) is more selective, a runtime filter (e.g., Bloom filter, bitmap) is generated over the join keys and injected into the scan operator of the document\_table. This early-stage pruning can both (1) improve the accuracy and (2) eliminate irrelevant rows. Otherwise, \sysname injects runtime filtering directly into operators like vector index scan, enabling coarse-grained pruning even within the vector retrieval phase.

\hi{(2) Fusion-based Hybrid Data Search.} With the \texttt{RANK\_FUSION} operator (used in Figure~\ref{fig:hybrid_query}), \sysname brings vector and text search together with relation query processing. Specifically, \texttt{RANK\_FUSION} is a specialized variant of the relational \textsf{Union} operator that offers two alternative fusion strategies. % This operator supports multiple fusion strategies for merging heterogeneous ranked lists.
The first is a {score-based fusion} approach, which performs Min–Max normalization within each modality to align score scales, followed by weighted linear aggregation of the normalized scores. Alternatively, \sysname employs a rank-based Reciprocal Rank Fusion (RRF) algorithm~\cite{cormack2009reciprocal}, which aggregates results purely based on their relative positions in ranked lists. Given $n$ ranked lists (e.g., from text and vector indexes), RRF assigns each document $d$ a unified relevance score: $RRF(d) = \sum^n_{i=1} \frac{1}{k+r_i(d)},$ where $r_i(d)$ denotes the rank position of document $d$ in the $i$-th modality-specific retrieval list, $k$ is a smoothing constant (typically set to $60$) that controls the decay of lower-ranked items, and $n$ is the number of participating modalities. By operating on rank positions rather than absolute similarity scores, RRF achieves modality-agnostic and calibration-free fusion.

\hi{(3) Selective Post-Join Refinement.} After the hybrid retriever produces the top-$K$ candidates, \sysname performs a selective post-join refinement to enforce structured predicates. Because runtime filtering (Step~(1)) has already pruned most irrelevant rows, this join operates on a substantially reduced candidate set.

\section{Performance Study}
\label{sec:exp}
%This section evaluates \sysname across representative workload scenarios. 
%We first evaluate baseline performance using standard analytical benchmarks, then analyze Incremental Processing Mode and \dc. 
%We then examine the impact of AI-driven optimization techniques, and conclude with an evaluation of hybrid multimodal query processing.

\subsection{Standard Benchmark Comparison}
We evaluate the analytical performance of ByteHouse operating against the latest StarRocks~\cite{starrocks}, Doris~\cite{doris}, and ClickHouse~\cite{schulze2024clickhouse} using two representative benchmarks: TPC-DS (SF-1000)~\cite{tpcds} and ClickBench~\cite{clickbench}.
All experiments are conducted on a dedicated cluster equipped with a 64-core AMD EPYC 9Y24 processor (128 hardware threads at 2.6 GHz) and 247 GB of main memory.
Figure~\ref{fig:exp_standard_benchmark} summarizes the comparative results.
On TPC-DS SF-1000 (Fig.~\ref{fig:subfig:tpcds}), \sysname consistently achieves lower end-to-end latency across all reported percentiles.
The performance gaps widen at higher percentiles, such as P95 and P99, where multi-stage aggregations and joins dominate execution time and amplify engine-level bottlenecks.
In these scenarios, \sysname achieves substantially lower tail latency than both StarRocks and Doris, indicating stable performance even under the most demanding queries.
Fig.~\ref{fig:subfig:clickbench} presents per-query results on the ClickBench benchmark, which consists of 43 queries executed five times each, with the fastest run recorded.
\sysname outperforms ClickHouse on approximately 32 queries and reduces the overall end-to-end latency by 25.4\%.
These performance gains are driven by \sysname's high-parallelism execution engine, adaptive aggregation and grouping mechanisms, optimized Top-N processing paths, and an efficient memory allocator that reduces allocation overhead and fragmentation under high concurrency.
%Together, these techniques minimize intermediate work and improve overall query efficiency.

\subsection{Effect of Incremental Processing}

\begin{figure}[h]
\hspace{-1em}
\centering
\subfloat[\small Inner Join Performance.]{
\label{fig:subfig:inner_join_tpch}
\vspace{-2ex}
{\includegraphics[scale=0.37]{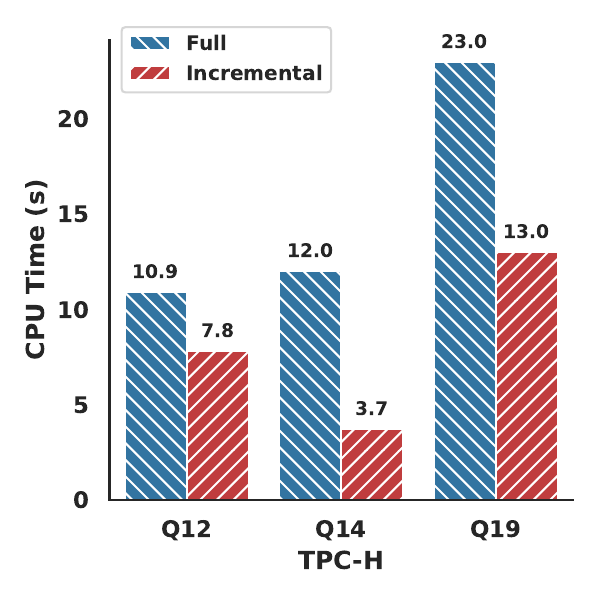}}
}
\hspace{-1em}
\subfloat[\small Impact of Update Ratio.]{
\label{fig:subfig:inner_join_incremental}
\vspace{-2ex}
{\includegraphics[scale=0.37]{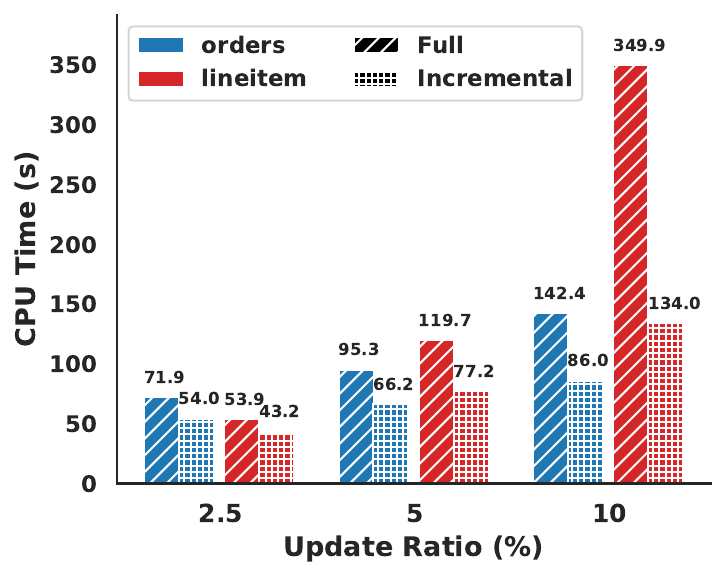}}
}
\vspace{-1em}
\caption{Efficiency of Incremental Processing.}
\vspace{-.5em}
\label{fig:exp_IPM}
\end{figure}

Figure~\ref{fig:exp_IPM} evaluates the efficiency of incremental processing mode (IPM) under representative inner join workloads from the TPC-H benchmark at scale factor 100.
Figure~\ref{fig:subfig:inner_join_tpch} compares the CPU time of full recomputation and incremental computation across three typical join queries (Q12, Q14, Q19), where updates are applied to the \textsf{lineitem} table with an update ratio of 2.5\%.
The incremental approach consistently outperforms the full recomputation baseline, reducing CPU time by 28.4\%–69.2\%, which highlights its advantage for processing complex multi-join plans.
Figure~\ref{fig:subfig:inner_join_incremental} further analyzes how incremental processing scales with different update ratios (2.5\%, 5\%, 10\%) using join-only SQL queries without any filter predicates, with updates applied to either \textsf{orders} or \textsf{lineitem}.
We measure the CPU time when applying updates to either the \textsf{orders} or \textsf{lineitem} table in query Q12.
As the update ratio increases from 2.5\% to 10\%, the CPU time reduction grows from 19.9\% up to 61.7\%, indicating that the incremental strategy becomes increasingly beneficial as the magnitude of data updates rises.
These results collectively demonstrate that incremental processing substantially improves computational efficiency, particularly in update-intensive analytical workloads.

\subsection{Effect of \dc}

\begin{figure}[h]
  \centering
  \includegraphics[scale=0.45]{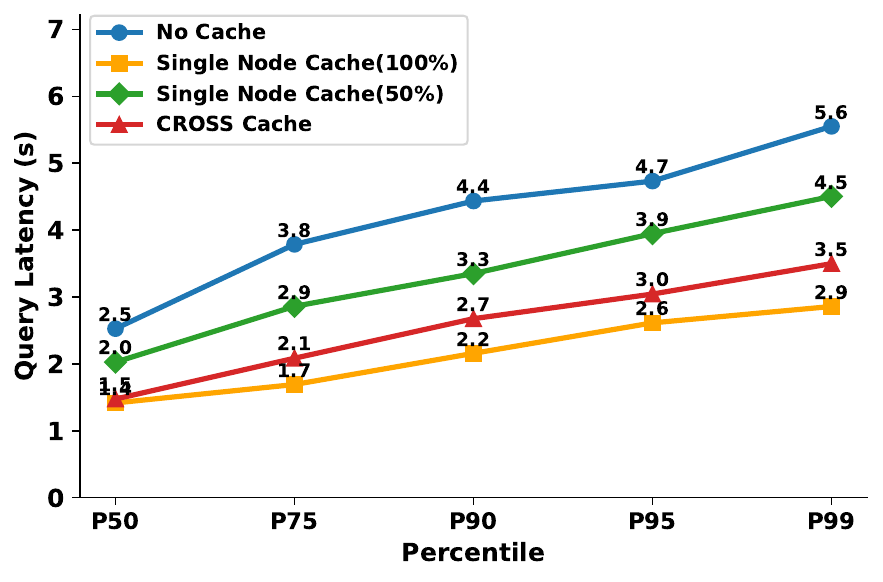}
  \vspace{-1.5em}
  \caption{Latency Evaluation of \dc.}
  \vspace{-1em}
  \label{fig:exp_cross}
\end{figure}

We evaluate the performance of \dc on a dedicated \sysname cluster used for large-scale advertising analytics, including user conversion analysis across ad impressions, clicks, and exposure events.
The cluster consists of 180 compute nodes, each equipped with 60 CPU cores (AMD EPYC 9Y24 96-Core Processor, 3.69 GHz), 256 GB of memory, and a 2.5 TB SSD.
\dc is deployed with one Cache Coordinator and 47 Cache Nodes, where each node provides 2.25 TB of local SSD cache.
The evaluated workload comprises the top $1000$ historical queries with the largest data scan volumes.
Each query completes within 5–7 s when served from cache, while full execution without caching typically takes around one minute.
To isolate the effect of distributed caching, we evaluate \dc and the single-node cache baselines in a fully isolated environment and enable only one caching mode per run to avoid cross-interference.

Figure~\ref{fig:exp_cross} reports end-to-end query latency percentiles (P50–P99) across four mutually exclusive settings: no cache, a single-node cache with a 100\% hit ratio, the same single-node cache with a 50\% hit ratio, and \dc.
The cache hit ratio is controlled by adjusting the available cache capacity.
Across all configurations, caching substantially reduces query latency.
Compared with the single-node cache at a 50\% hit ratio, \dc reduces latency across all percentiles, with improvements of about 25\% at P50, 18\% at P90, and 22\% at P99.
While \dc exhibits marginally higher latency than the idealized single-node cache with a perfect (100\%) hit ratio, it consistently achieves lower latency than the single-node cache at a 50\% hit ratio.
This advantage is particularly meaningful in production environments, where maintaining perfect cache locality on a single node is rarely achievable.
%Overall, \dc delivers near-local cache performance while offering stronger consistency, scalability, and robustness under realistic cache conditions.

\subsection{Effect of AI-driven Optimization}

\begin{figure}[h]
\hspace{-1em}
\centering
\subfloat[\small Read Volume Reduction.]{
\label{fig:subfig:prewhere_read_bytes}
{\includegraphics[scale=0.38]{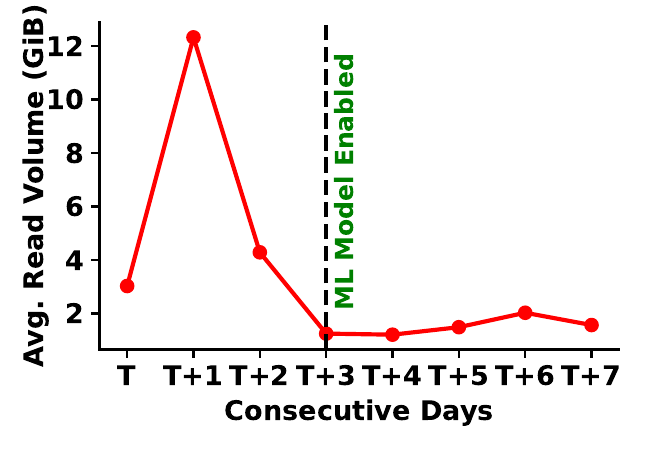}}
}
\hspace{-1em}
\subfloat[\small Timeout Rate ($\mathbf{\ge 10}$ s).]{
\label{fig:subfig:prewhere_timeout}
{\includegraphics[scale=0.38]{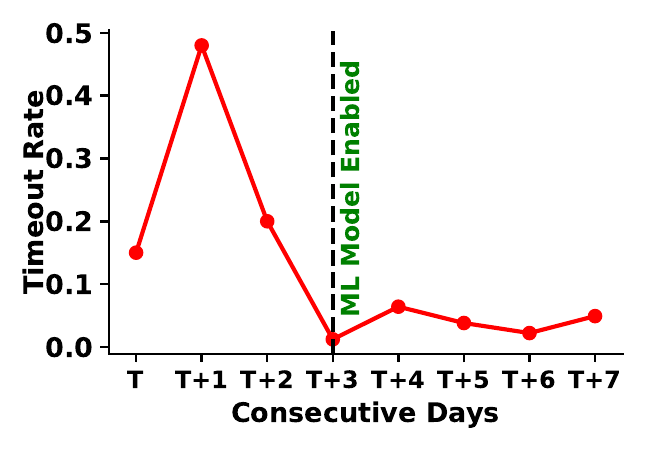}}
}
\ \ \\
\subfloat[\small Query Latency Reduction.]{
\label{fig:subfig:prewhere_latency}
\vspace{-2ex}
{\includegraphics[scale=0.38]{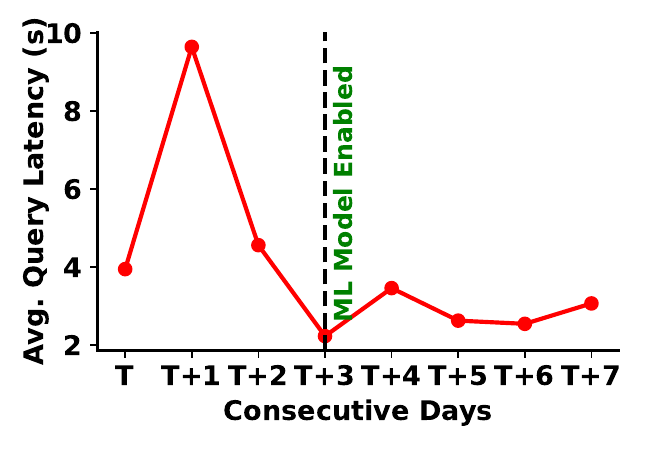}}
}
\hspace{-1em}
\subfloat[\small AI4JSS Performance.]{
\label{fig:subfig:exp_joinreorder}
\vspace{-2ex}
{\includegraphics[scale=0.36]{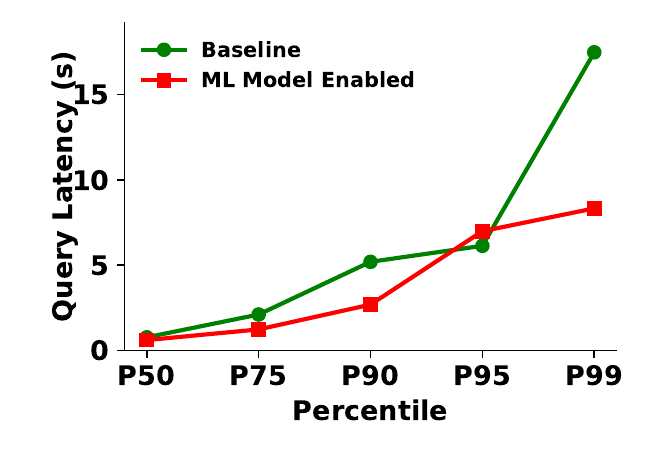}}
}
\vspace{-1em}
\caption{Efficiency of AI-driven Optimization.}\
\vspace{-1em}
\label{fig:exp_AI}
\end{figure}

Figure~\ref{fig:exp_AI} illustrates the performance  of two AI-driven approaches on query execution efficiency.
The experiment was conducted on a production cluster consisting of 100 nodes, each equipped with 192 CPU cores and 1.48~TiB of memory.
Figures~\ref{fig:subfig:prewhere_read_bytes}-\ref{fig:subfig:prewhere_latency} present the results of the Predicate Pushdown Selectivity (PPS) experiment, where an ML model predicts filter selectivity and dynamically adjusts predicate pushdown strategies.
After enabling the model (day T+3), the system exhibits notable reductions in both I/O and query latency: the average read volume drops by up to 87\%, the proportion of queries exceeding 10 seconds (timeout rate) drops sharply, and the average query latency is reduced accordingly.
Figure~\ref{fig:subfig:exp_joinreorder} further evaluates the Join Side Selection (JSS) experiment, where an ML model is employed to optimize build/probe side selection.
The experiment evaluates 1,000 join queries sampled from real production traces, comparing the latency distributions of the proposed approach against the baseline optimizer.
Across all percentiles, the AI-driven approach achieves 15\%–45\% lower query latency, with particularly strong improvements at the tail (P95–P99) where long-running queries dominate overall cost.
Across all latency percentiles (P50–P99), the ML model consistently outperforms the baseline, with the most significant gains observed in the tail latencies (P95–P99).
Overall, these results confirm that AI-driven optimization substantially improves scan and join efficiency, delivering measurable gains in real-world analytical query processing.

\subsection{Hybrid Multimodal Query Processing}

\begin{figure}[h]
\hspace{-1em}
\centering
\subfloat[\small Vector Retrieval Throughput.]{
\label{fig:subfig:exp_vector}
\vspace{-2ex}
{\includegraphics[scale=0.33]{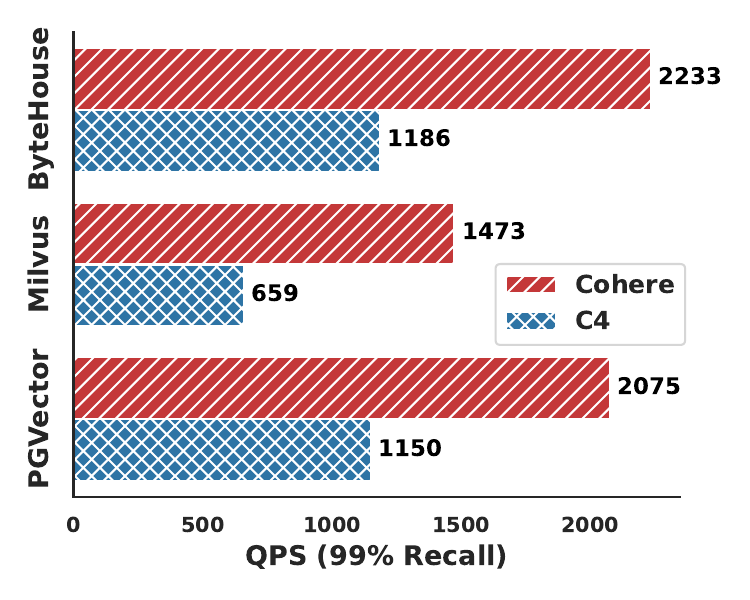}}
}
\hspace{-1em}
\subfloat[\small Hybrid Search Recall.]{
\label{fig:subfig:exp_hybrid}
{\includegraphics[scale=0.33]{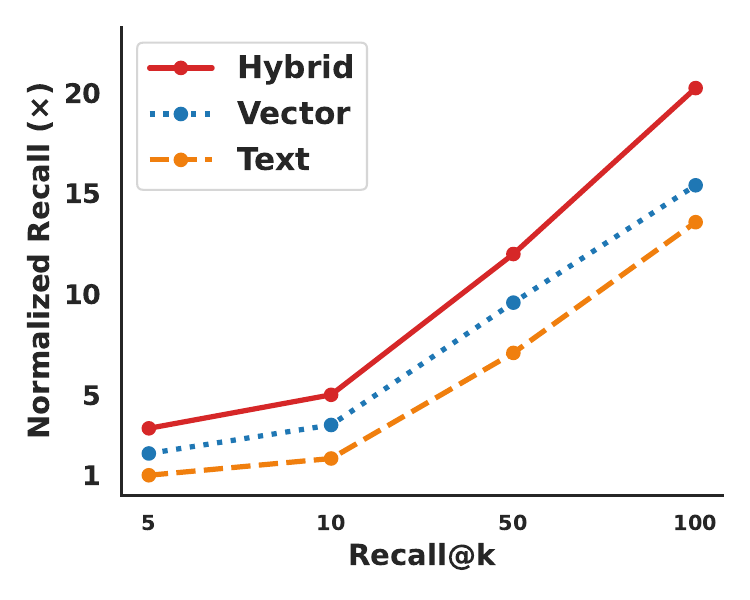}}
}
\vspace{-.5em}
\caption{Performance of Multimodal Query Processing.}
%\vspace{-.5em}
\label{fig:exp_multimodal}
\end{figure}

Figure~\ref{fig:exp_multimodal} evaluates the multimodal query processing capability of \sysname.
Figure~\ref{fig:subfig:exp_vector} reports vector retrieval throughput (QPS at 99 \% recall) on the Cohere~\cite{cohere_wikipedia_22_12} and C4~\cite{allenai_c4} datasets using the VDBBench framework~\cite{VectorDBBench}, comparing \sysname with Milvus~\cite{wang2021milvus} and pgvector~\cite{pgvector}.
The workload consists of hybrid queries that combine vector similarity search with a 1\% scalar filter.
Across both datasets, \sysname achieves the highest throughput, improving performance by 50–60\% over Milvus on Cohere and by more than 50\% on C4, while also maintaining a clear lead over pgvector.
These gains stem from \sysname's multi-layer vector index choices and cross-table runtime filtering.

Figure~\ref{fig:subfig:exp_hybrid} presents the multimodal retrieval accuracy using the MS MARCO~\cite{bajaj2016ms}, which contains 8.84M passages (we sample 2M for testing).
Each passage is represented by the BGE-M3 embedding model~\cite{bge-m3}.
We evaluate three retrieval approaches in \sysname (i.e., Vector Search, Text search, and Hybrid Search) under the recall metric at different top-$k$ thresholds.
The results shows that Hybrid Search consistently achieves the best overall recall, outperforming the single-modality baselines.
At Recall@100, it yields approximately a 30\% improvement over vector search and a 50\% improvement over text search. These results demonstrate that \sysname effectively integrates semantic embedding signals with lexical matching within a unified execution framework.

\section{Lessons Learned}

\hi{Durability and Data Organization.}
Our experience indicates that
storage latency is dominated by the write path, where durability and isolation guarantees must be enforced during ingestion.
Write-through policies~\cite{verbitski2017amazon,corbett2013spanner} simplify reliability but often suffer from high cold-read latency under cache misses, whereas write-back policies~\cite{merli2025ursa,antonopoulos2019socrates} hide write latency through buffering and deferred persistence, which increases the complexity of durability semantics. 
A key lesson across these policies is that durability and availability should be decoupled so that persistence can proceed asynchronously without compromising service continuity.
We also believe that hybrid data layouts~\cite{schmidt2024two} that keep hot tuples in a row format and colder regions in a compressed columnar format provide a practical way to accommodate both update-intensive operations and bandwidth-efficient analytical scans.

\hi{Sustaining High Concurrency.}
\sysname serves analytical workloads that reach $10^4-10^5$ read/write QPS, where per-query control overheads, metadata lock contention, and I/O amplification from small fragmented accesses quickly become bottlenecks.
Our experience indicates that achieving stable performance at this scale requires coordinated design across planning, compute, and storage, rather than isolated operator-level optimizations.
\sysname reduces planning-layer control overheads through \textit{prepared statements}, accelerates execution-layer performance for selective workloads (e.g., TopN) via \textit{short-circuit evaluation},
and employs the \dc layer to absorb fine-grained accesses and reduce storage-layer I/O amplification.
This end-to-end co-design enables scalable throughput with stable tail latency at high concurrency.

\hi{Vector Search and Multi-Layer Indexing.} 
A key lesson is that vector computation is becoming a fundamental building block for multimodal analytics.
Similarity search workloads are moving beyond simple Top-$k$ retrieval with scalar filters toward richer hybrid patterns, such as similarity-bounded range aggregation~\cite{lan2024cardinality,liang2024unify}, vector-driven word-cloud analysis~\cite{wu2011semantic,castella2014word,schubert2017semantic}, and multi-vector joint recall across embedding spaces~\cite{chen2024m3,li2023slim}.
These trends highlight the need for more flexible vector-query semantics and tighter integration of vector search with traditional analytical operators to support increasingly expressive multimodal workloads.

\vspace{.5em}
\section{Conclusions}
\vspace{.5em}

\begin{sloppypar}

This paper presents \sysname, a cloud-native shared-storage data warehouse for real-time multimodal analytics. 
\sysname integrates a vertically optimized storage layer with a unified execution framework that supports analytic, batch, and incremental processing. Its fusion-based retrieval operators and hybrid optimization techniques enable efficient multimodal query execution.
Evaluations across production and standard benchmarks confirm that \sysname is a scalable, high-performance foundation for emerging intelligent data applications.
\end{sloppypar}

\begin{acks}
\begin{sloppypar}
Xuanhe Zhou is the corresponding author.
We thank Tao Wang and Xuechao Lian for contributions to the development of SBM and IPM, Lixiang Qian for APM, Chengxian Li for traditional optimizer, 
 Changheng Cai for hybrid data search, 
Zichun Huang for \fs, and Xuewu Xiao, Qiang Ren, and Shuo Jia for \dc.
This work was supported in part by National Key R\&D Program of China (No. 2023YFB4502400), 
%China NSF grant (No. 62441236, 62372296, 62432007, U25A6024, 62502304, U25A20437), 
%Fundamental and Interdisciplinary Disciplines Breakthrough Plan of the Ministry of Education of China (No. JYB2025XDXM103), ByteDance, 
Shanghai Jiao Tong University AI for Engineering Initiative and Shanghai Qi Zhi Institute Innovation Program.
\end{sloppypar}
\end{acks}

\clearpage
\balance
\bibliographystyle{ACM-Reference-Format}
\bibliography{ref}

@inproceedings{zhou2010incorporating,
  title={Incorporating partitioning and parallel plans into the SCOPE optimizer},
  author={Zhou, Jingren and Larson, Per-Ake and Chaiken, Ronnie},
  booktitle={2010 IEEE 26th International Conference on Data Engineering (ICDE 2010)},
  pages={1060--1071},
  year={2010},
  organization={IEEE}
}

@inproceedings{yang2023fifo,
  title={FIFO queues are all you need for cache eviction},
  author={Yang, Juncheng and Zhang, Yazhuo and Qiu, Ziyue and Yue, Yao and Vinayak, Rashmi},
  booktitle={Proceedings of the 29th Symposium on Operating Systems Principles},
  pages={130--149},
  year={2023}
}

@inproceedings{Zhang2024PhatKV,
  author       = {Yiwen Zhang and Guokuan Li and Kai Lu and Jiguang Wan and Ting Yao and Huatao Wu and Daohui Wang},
  title        = {{PhatKV}: Towards an Efficient Metadata Engine for KV-based File Systems on Modern SSD},
  booktitle    = {Proceedings of the MSST 2024},
  year         = {2024},
  organization = {IEEE},
}

@article{kuschewski2023btrblocks,
  title={Btrblocks: Efficient columnar compression for data lakes},
  author={Kuschewski, Maximilian and Sauerwein, David and Alhomssi, Adnan and Leis, Viktor},
  journal={SIGMOD},
  volume={1},
  number={2},
  pages={1--26},
  year={2023},
  publisher={ACM New York, NY, USA}
}

@unpublished{zeng2026f3,
  author       = {Xinyu Zeng and Ruijun Meng and Martin Prammer and Wes McKinney and Jignesh M. Patel and Andrew Pavlo and Huanchen Zhang},
  title        = {F3: The Open-Source Data File Format for the Future},
  note         = {SIGMOD)},
  year         = {2026},
}

@article{afroozeh2023alp,
  title={{ALP}: Adaptive lossless floating-point compression},
  author={Afroozeh, Azim and Kuffo, Leonardo X and Boncz, Peter},
  journal={SIGMOD},
  volume={1},
  number={4},
  pages={1--26},
  year={2023},
}

@inproceedings{niu2025blendhouse,
  title={{BlendHouse}: A Cloud-Native Vector Database System in ByteHouse},
  author={Niu, Zhaojie and Tian, Xinhui and Peng, Xindong and Chen, Xing},
  booktitle={2025 IEEE 41st International Conference on Data Engineering (ICDE)},
  pages={4332--4345},
  year={2025},
  organization={IEEE}
}

@article{graefe1995cascades,
  title={The cascades framework for query optimization},
  author={Graefe, Goetz},
  journal={IEEE Data Eng. Bull.},
  volume={18},
  number={3},
  pages={19--29},
  year={1995}
}

@article{shankhdhar2024presto,
  title={Presto's History-Based Query Optimizer},
  author={Shankhdhar, Pranjal and Liu, Feilong and Narale, Jay and Sun, James and Schlussel, Rebecca and Antova, Lyublena},
  journal={Proceedings of the VLDB Endowment},
  volume={17},
  number={12},
  pages={4077--4089},
  year={2024},
  publisher={VLDB Endowment}
}

@inproceedings{berg2020cachelib,
  title={The $\{$CacheLib$\}$ caching engine: Design and experiences at scale},
  author={Berg, Benjamin and Berger, Daniel S and McAllister, Sara and Grosof, Isaac and Gunasekar, Sathya and Lu, Jimmy and Uhlar, Michael and Carrig, Jim and Beckmann, Nathan and Harchol-Balter, Mor and others},
  booktitle={14th USENIX Symposium on Operating Systems Design and Implementation (OSDI 20)},
  pages={753--768},
  year={2020}
}

@article{budiu2022dbsp,
    author = {Budiu, Mihai and Chajed, Tej and McSherry, Frank and Ryzhyk, Leonid and Tannen, Val},
    title = {{DBSP}: Automatic Incremental View Maintenance for Rich Query Languages},
    year = {2023},
    volume = {16},
    number = {7},
    journal = {Proc. VLDB Endow.},
    pages = {1601–1614},
    numpages = {14}
}

@misc{apachearrow,
  title        = {Apache Arrow: A cross-language development platform for in-memory data},
  author       = {{The Apache Software Foundation}},
  year         = {2025},
  howpublished = {\url{https://arrow.apache.org/}},
  note         = {Accessed: 2025-10-24}
}

@article{schulze2024clickhouse,
  title={Clickhouse-lightning fast analytics for everyone},
  author={Schulze, Robert and Schreiber, Tom and Yatsishin, Ilya and Dahimene, Ryadh and Milovidov, Alexey},
  journal={Proceedings of the VLDB Endowment},
  volume={17},
  number={12},
  pages={3731--3744},
  year={2024},
}

@inproceedings{han2024bytecard,
  title={ByteCard: Enhancing ByteDance's Data Warehouse with Learned Cardinality Estimation},
  author={Han, Yuxing and Wang, Haoyu and Chen, Lixiang and Dong, Yifeng and Chen, Xing and Yu, Benquan and Yang, Chengcheng and Qian, Weining},
  booktitle={Companion of the 2024 International Conference on Management of Data},
  pages={41--54},
  year={2024}
}

@article{cte-optimization,
  title={Optimization of common table expressions in mpp database systems},
  author={El-Helw, Amr and Raghavan, Venkatesh and Soliman, Mohamed A and Caragea, George and Gu, Zhongxian and Petropoulos, Michalis},
  journal={VLDB},
  volume={8},
  number={12},
  pages={1704--1715},
  year={2015},
}

@inproceedings{magicset,
  title={Cost-based optimization for magic: Algebra and implementation},
  author={Seshadri, Praveen and Hellerstein, Joseph M and Pirahesh, Hamid and Leung, TY Cliff and Ramakrishnan, Raghu and Srivastava, Divesh and Stuckey, Peter J and Sudarshan, S},
  booktitle={SIGMOD},
  pages={435--446},
  year={1996}
}

@inproceedings{saxena2023auto,
  title={{Auto-WLM}: Machine learning enhanced workload management in Amazon Redshift},
  author={Saxena, Gaurav and Rahman, Mohammad and Chainani, Naresh and Lin, Chunbin and Caragea, George and Chowdhury, Fahim and Marcus, Ryan and Kraska, Tim and Pandis, Ippokratis and Narayanaswamy, Balakrishnan},
  booktitle={Companion of the 2023 International Conference on Management of Data},
  pages={225--237},
  year={2023}
}

@inproceedings{graefe1993volcano,
  title={The volcano optimizer generator: Extensibility and efficient search},
  author={Graefe, Goetz and McKenna, William J},
  booktitle={Proceedings of IEEE 9th international conference on data engineering},
  pages={209--218},
  year={1993},
  organization={IEEE}
}

@inproceedings{soliman2014orca,
  title={{Orca}: a modular query optimizer architecture for big data},
  author={Soliman, Mohamed A and Antova, Lyublena and Raghavan, Venkatesh and El-Helw, Amr and Gu, Zhongxian and Shen, Entong and Caragea, George C and Garcia-Alvarado, Carlos and Rahman, Foyzur and Petropoulos, Michalis and others},
  booktitle={Proceedings of the 2014 ACM SIGMOD international conference on Management of data},
  pages={337--348},
  year={2014}
}

@inproceedings{silva2012exploiting,
  title={Exploiting common subexpressions for cloud query processing},
  author={Silva, Yasin N and Larson, Paul-Ake and Zhou, Jingren},
  booktitle={2012 IEEE 28th International Conference on Data Engineering},
  pages={1337--1348},
  year={2012},
  organization={IEEE}
}

@inproceedings{sereni2008adding,
  title={Adding magic to an optimising datalog compiler},
  author={Sereni, Damien and Avgustinov, Pavel and De Moor, Oege},
  booktitle={Proceedings of the 2008 ACM SIGMOD international conference on Management of data},
  pages={553--566},
  year={2008}
}

@inproceedings{galindo2003statistics,
  title={Statistics on views},
  author={Galindo-Legaria, C{\'e}sar A and Joshi, Milind M and Waas, Florian and Wu, Ming-Chuan},
  booktitle={Proceedings 2003 VLDB Conference},
  pages={952--962},
  year={2003},
  organization={Elsevier}
}

@article{neo,
author = {Marcus, Ryan and Negi, Parimarjan and Mao, Hongzi and Zhang, Chi and Alizadeh, Mohammad and Kraska, Tim and Papaemmanouil, Olga and Tatbul, Nesime},
title = {Neo: a learned query optimizer},
year = {2019},
volume = {12},
number = {11},
journal = {Proc. VLDB Endow.},
month = jul,
pages = {1705–1718},
numpages = {14}
}

@article{zhu2020flat,
  title={{FLAT}: Fast, Lightweight and Accurate Method for Cardinality Estimation},
  author={Zhu, Rong and Wu, Ziniu and Han, Yuxing and Zeng, Kai and Pfadler, Andreas and Qian, Zhengping and Zhou, Jingren and Cui, Bin},
  journal={VLDB},
  pages={1489--1502},
  volume={14},
  number={9},
  year={2021}
}

@article{SunL19,
  author       = {Ji Sun and
                  Guoliang Li},
  title        = {An End-to-End Learning-based Cost Estimator},
  journal      = {Proc. {VLDB} Endow.},
  volume       = {13},
  number       = {3},
  pages        = {307--319},
  year         = {2019},
}

@book{koller2009probabilistic,
  title={Probabilistic graphical models: principles and techniques},
  author={Koller, Daphne and Friedman, Nir},
  year={2009},
  publisher={MIT press}
}

@inproceedings{fender2011femo,
  author       = {Pit Fender and
                  Guido Moerkotte},
  editor       = {Serge Abiteboul and
                  Klemens B{\"{o}}hm and
                  Christoph Koch and
                  Kian{-}Lee Tan},
  title        = {A new, highly efficient, and easy to implement top-down join enumeration
                  algorithm},
  booktitle    = {ICDE},
  pages        = {864--875},
  year         = {2011}
}

@article{akidau2023s,
      title={What's the Difference? Incremental Processing with Change Queries in Snowflake},
      author={Akidau, Tyler and Barbier, Paul and Cseri, Istvan and Hueske, Fabian and Jones, Tyler and Lionheart, Sasha and Mills, Daniel and Pauliukevich, Dzmitry and Probst, Lukas and Semmler, Niklas and others},
      journal={Proceedings of the ACM on Management of Data},
      volume={1},
      number={2},
      pages={1--27},
      year={2023},
}

@inproceedings{sotolongo2025streaming,
  title={Streaming Democratized: Ease Across the Latency Spectrum with Delayed View Semantics and Snowflake Dynamic Tables},
  author={Sotolongo, Daniel and Mills, Daniel and Akidau, Tyler and Santhiar, Anirudh and T{\'o}th, Attila-P{\'e}ter and Huang, Botong and Zhang, Boyuan and Belianski, Igor and Geng, Ling and Uhlar, Matt and others},
  booktitle={Companion of the 2025 International Conference on Management of Data},
  pages={622--634},
  year={2025}
}

@article{wang2020tempura,
  title={{Tempura}: a general cost-based optimizer framework for incremental data processing},
  author={Wang, Zuozhi and Zeng, Kai and Huang, Botong and Chen, Wei and Cui, Xiaozong and Wang, Bo and Liu, Ji and Fan, Liya and Qu, Dachuan and Hou, Zhenyu and others},
  journal={Proceedings of the VLDB Endowment},
  volume={14},
  number={1},
  pages={14--27},
  year={2020},
  publisher={VLDB Endowment}
}

@article{gao2025trae,
  title={Trae agent: An llm-based agent for software engineering with test-time scaling},
  author={Gao, Pengfei and Tian, Zhao and Meng, Xiangxin and Wang, Xinchen and Hu, Ruida and Xiao, Yuanan and Liu, Yizhou and Zhang, Zhao and Chen, Junjie and Gao, Cuiyun and others},
  journal={arXiv preprint arXiv:2507.23370},
  year={2025}
}

@article{merli2025ursa,
  title={{Ursa}: A Lakehouse-Native Data Streaming Engine for Kafka},
  author={Merli, Matteo and Guo, Sijie and Li, Penghui and Chen, Hang and Lu, Neng},
  journal={Proceedings of the VLDB Endowment},
  volume={18},
  number={12},
  pages={5184--5196},
  year={2025},
  publisher={VLDB Endowment}
}

@inproceedings{antonopoulos2019socrates,
  title={Socrates: The new sql server in the cloud},
  author={Antonopoulos, Panagiotis and Budovski, Alex and Diaconu, Cristian and Hernandez Saenz, Alejandro and Hu, Jack and Kodavalla, Hanuma and Kossmann, Donald and Lingam, Sandeep and Minhas, Umar Farooq and Prakash, Naveen and others},
  booktitle={Proceedings of the 2019 International Conference on Management of Data},
  pages={1743--1756},
  year={2019}
}

@article{schmidt2024two,
  title={Two Birds With One Stone: Designing a Hybrid Cloud Storage Engine for HTAP},
  author={Schmidt, Tobias and Durner, Dominik and Leis, Viktor and Neumann, Thomas},
  journal={Proceedings of the VLDB Endowment},
  volume={17},
  number={11},
  pages={3290--3303},
  year={2024},
  publisher={VLDB Endowment}
}

@inproceedings{verbitski2017amazon,
  title={Amazon {Aurora}: Design considerations for high throughput cloud-native relational databases},
  author={Verbitski, Alexandre and Gupta, Anurag and Saha, Debanjan and Brahmadesam, Murali and Gupta, Kamal and Mittal, Raman and Krishnamurthy, Sailesh and Maurice, Sandor and Kharatishvili, Tengiz and Bao, Xiaofeng},
  booktitle={Proceedings of the 2017 ACM International Conference on Management of Data},
  pages={1041--1052},
  year={2017}
}

@article{corbett2013spanner,
  title={Spanner: Google’s globally distributed database},
  author={Corbett, James C and Dean, Jeffrey and Epstein, Michael and Fikes, Andrew and Frost, Christopher and Furman, Jeffrey John and Ghemawat, Sanjay and Gubarev, Andrey and Heiser, Christopher and Hochschild, Peter and others},
  journal={ACM Transactions on Computer Systems (TOCS)},
  volume={31},
  number={3},
  pages={1--22},
  year={2013},
  publisher={ACM New York, NY, USA}
}

@inproceedings{wu2011semantic,
  title={Semantic-preserving word clouds by seam carving},
  author={Wu, Yingcai and Provan, Thomas and Wei, Furu and Liu, Shixia and Ma, Kwan-Liu},
  booktitle={Computer Graphics Forum},
  volume={30},
  number={3},
  pages={741--750},
  year={2011},
  organization={Wiley Online Library}
}

@inproceedings{castella2014word,
  title={Word storms: Multiples of word clouds for visual comparison of documents},
  author={Castella, Quim and Sutton, Charles},
  booktitle={Proceedings of the 23rd international conference on World wide web},
  pages={665--676},
  year={2014}
}

@article{schubert2017semantic,
  title={Semantic word clouds with background corpus normalization and t-distributed stochastic neighbor embedding},
  author={Schubert, Erich and Spitz, Andreas and Weiler, Michael and Gei{\ss}, Johanna and Gertz, Michael},
  journal={arXiv preprint arXiv:1708.03569},
  year={2017}
}

@article{lan2024cardinality,
  title={Cardinality Estimation for Similarity Search on High-Dimensional Data Objects: The Impact of Reference Objects},
  author={Lan, Hai and Huang, Shixun and Bao, Zhifeng and Borovica-Gajic, Renata},
  journal={Proceedings of the VLDB Endowment},
  volume={18},
  number={3},
  pages={544--556},
  year={2024},
  publisher={VLDB Endowment}
}

@article{liang2024unify,
  title={UNIFY: Unified Index for Range Filtered Approximate Nearest Neighbors Search},
  author={Liang, Anqi and Zhang, Pengcheng and Yao, Bin and Chen, Zhongpu and Song, Yitong and Cheng, Guangxu},
  journal={arXiv preprint arXiv:2412.02448},
  year={2024}
}

@inproceedings{chen2024m3,
  title={M3-embedding: Multi-linguality, multi-functionality, multi-granularity text embeddings through self-knowledge distillation},
  author={Chen, Jianlyu and Xiao, Shitao and Zhang, Peitian and Luo, Kun and Lian, Defu and Liu, Zheng},
  booktitle={Findings of the Association for Computational Linguistics ACL 2024},
  pages={2318--2335},
  year={2024}
}

@inproceedings{li2023slim,
  title={Slim: Sparsified late interaction for multi-vector retrieval with inverted indexes},
  author={Li, Minghan and Lin, Sheng-Chieh and Ma, Xueguang and Lin, Jimmy},
  booktitle={Proceedings of the 46th International ACM SIGIR Conference on Research and Development in Information Retrieval},
  pages={1954--1959},
  year={2023}
}

@inproceedings{cormack2009reciprocal,
  title={Reciprocal rank fusion outperforms condorcet and individual rank learning methods},
  author={Cormack, Gordon V and Clarke, Charles LA and Buettcher, Stefan},
  booktitle={Proceedings of the 32nd international ACM SIGIR conference on Research and development in information retrieval},
  pages={758--759},
  year={2009}
}

@article{malkov2018hnsw,
  title={Efficient and robust approximate nearest neighbor search using hierarchical navigable small world graphs},
  author={Malkov, Yu A and Yashunin, Dmitry A},
  journal={IEEE transactions on pattern analysis and machine intelligence},
  volume={42},
  number={4},
  pages={824--836},
  year={2018},
  publisher={IEEE}
}

@article{jayaram2019diskann,
  title={{DiskAnn}: Fast accurate billion-point nearest neighbor search on a single node},
  author={Jayaram Subramanya, Suhas and Devvrit, Fnu and Simhadri, Harsha Vardhan and Krishnawamy, Ravishankar and Kadekodi, Rohan},
  journal={Advances in neural information processing Systems},
  volume={32},
  year={2019}
}

@inproceedings{MSCN,
	title={Learned cardinalities: Estimating correlated joins with deep learning},
	author={Kipf, Andreas and Kipf, Thomas and Radke, Bernhard and Leis, Viktor and Boncz, Peter and Kemper, Alfons},
	booktitle={CIDR},
	year={2019}
}

@inproceedings{jeong2025upp,
  title={{UPP}: Universal Predicate Pushdown to Smart Storage},
  author={Jeong, Ipoom and Huang, Jinghan and Hu, Chuxuan and Park, Dohyun and Kang, Jaeyoung and Kim, Nam Sung and Park, Yongjoo},
  booktitle={Proceedings of the 52nd Annual International Symposium on Computer Architecture},
  pages={419--433},
  year={2025}
}

@article{yan2023predicate,
  title={Predicate pushdown for data science pipelines},
  author={Yan, Cong and Lin, Yin and He, Yeye},
  journal={Proceedings of the ACM on Management of Data},
  volume={1},
  number={2},
  pages={1--28},
  year={2023},
  publisher={ACM New York, NY, USA}
}

@article{wu2020bayescard,
  title={Bayescard: Revitilizing bayesian frameworks for cardinality estimation},
  author={Wu, Ziniu and Shaikhha, Amir and Zhu, Rong and Zeng, Kai and Han, Yuxing and Zhou, Jingren},
  journal={arXiv preprint arXiv:2012.14743},
  year={2020}
}

@article{gholamalinezhad2020pooling,
  title={Pooling methods in deep neural networks, a review},
  author={Gholamalinezhad, Hossein and Khosravi, Hossein},
  journal={arXiv preprint arXiv:2009.07485},
  year={2020}
}

@article{MarcusNMZAKPT19,
  author       = {Ryan Marcus and
                  Parimarjan Negi and
                  Hongzi Mao and
                  Chi Zhang and
                  Mohammad Alizadeh and
                  Tim Kraska and
                  Olga Papaemmanouil and
                  Nesime Tatbul},
  title        = {Neo: {A} Learned Query Optimizer},
  journal      = {Proc. {VLDB} Endow.},
  volume       = {12},
  number       = {11},
  pages        = {1705--1718},
  year         = {2019},
}

@inproceedings{trummer2019skinnerdb,
  title={Skinnerdb: Regret-bounded query evaluation via reinforcement learning},
  author={Trummer, Immanuel and Wang, Junxiong and Maram, Deepak and Moseley, Samuel and Jo, Saehan and Antonakakis, Joseph},
  booktitle={Proceedings of the 2019 International Conference on Management of Data},
  pages={1153--1170},
  year={2019}
}

@article{ZhuCDCPWZ23,
  author       = {Rong Zhu and
                  Wei Chen and
                  Bolin Ding and
                  Xingguang Chen and
                  Andreas Pfadler and
                  Ziniu Wu and
                  Jingren Zhou},
  title        = {Lero: {A} Learning-to-Rank Query Optimizer},
  journal      = {Proc. {VLDB} Endow.},
  volume       = {16},
  number       = {6},
  pages        = {1466--1479},
  year         = {2023},
}

@inproceedings{snowflake,
  author       = {Beno{\^{\i}}t Dageville and
                  Thierry Cruanes and
                  Marcin Zukowski and
                  Vadim Antonov and
                  Artin Avanes and
                  Jon Bock and
                  Jonathan Claybaugh and
                  Daniel Engovatov and
                  Martin Hentschel and
                  Jiansheng Huang and
                  Allison W. Lee and
                  Ashish Motivala and
                  Abdul Q. Munir and
                  Steven Pelley and
                  Peter Povinec and
                  Greg Rahn and
                  Spyridon Triantafyllis and
                  Philipp Unterbrunner},
  title        = {The Snowflake Elastic Data Warehouse},
  booktitle    = {{SIGMOD} Conference},
  pages        = {215--226},
  publisher    = {{ACM}},
  year         = {2016}
}

@misc{redshift,
  title        = {Amazon Redshift},
  howpublished = {\url{https://aws.amazon.com/redshift/}},
  year         = {2025},
  note         = {Accessed: 2025-10-27}
}

@misc{clickbench,
  title        = {ClickBench: A Benchmark for Analytical DBMS},
  author       = {ClickHouse Team},
  howpublished = {\url{https://github.com/ClickHouse/ClickBench}},
  note         = {Accessed: 2025-10-27},
  year         = {2021}
}

@misc{tpcds,
  author       = {{Transaction Processing Performance Council}},
  title        = {TPC-DS Benchmark (Version 3.2)},
  year         = {2018},
  howpublished = {\url{http://www.tpc.org/tpcds/}},
  note         = {Accessed: 2025-10-27}
}

@misc{starrocks,
  author       = {{StarRocks Team}},
  title        = {StarRocks: A High-Performance MPP Database for Analytics},
  year         = {2020},
  howpublished = {\url{https://github.com/StarRocks/starrocks}},
  note         = {Accessed: 2025-10-27}
}

@misc{doris,
  author       = {{Apache Doris Community}},
  title        = {{Apache Doris}: An MPP Analytical Database for Real-Time Analytics},
  year         = {2017},
  howpublished = {\url{https://github.com/apache/doris}},
  note         = {Accessed: 2025-10-27}
}

@article{GaussDB,
author = {Li, Guoliang and Tian, Wengang and Zhang, Jinyu and Grosman, Ronen and Liu, Zongchao and Li, Sihao},
title = {GaussDB: A Cloud-Native Multi-Primary Database with Compute-Memory-Storage Disaggregation},
year = {2024},
publisher = {VLDB Endowment},
volume = {17},
number = {12},
journal = {Proc. VLDB Endow.},
pages = {3786–3798},
numpages = {13}
}

@inproceedings{PolardbMP,
author = {Yang, Xinjun and Zhang, Yingqiang and Chen, Hao and Li, Feifei and Wang, Bo and Fang, Jing and Sun, Chuan and Wang, Yuhui},
title = {PolarDB-MP: A Multi-Primary Cloud-Native Database via Disaggregated Shared Memory},
year = {2024},
booktitle = {Companion of the 2024 International Conference on Management of Data},
pages = {295–308},
numpages = {14},
}

@article{hologres,
  author       = {Xiaowei Jiang and
                  Yuejun Hu and
                  Yu Xiang and
                  Guangran Jiang and
                  Xiaojun Jin and
                  Chen Xia and
                  Weihua Jiang and
                  Jun Yu and
                  Haitao Wang and
                  Yuan Jiang and
                  Jihong Ma and
                  Li Su and
                  Kai Zeng},
  title        = {Alibaba Hologres: {A} Cloud-Native Service for Hybrid Serving/Analytical
                  Processing},
  journal      = {Proc. {VLDB} Endow.},
  volume       = {13},
  number       = {12},
  pages        = {3272--3284},
  year         = {2020},
}

@article{AnalyticDB,
  author       = {Chaoqun Zhan and
                  Maomeng Su and
                  Chuangxian Wei and
                  Xiaoqiang Peng and
                  Liang Lin and
                  Sheng Wang and
                  Zhe Chen and
                  Feifei Li and
                  Yue Pan and
                  Fang Zheng and
                  Chengliang Chai},
  title        = {AnalyticDB: Real-time {OLAP} Database System at Alibaba Cloud},
  journal      = {Proc. {VLDB} Endow.},
  volume       = {12},
  number       = {12},
  pages        = {2059--2070},
  year         = {2019}
}

@misc{vortex2024,
  title        = {Vortex: A Next-Generation High-Performance Data Format for AI and Analytics},
  author       = {{Vortex Project Team}},
  year         = {2024},
  howpublished = {\url{https://vortex.dev}},
  institution  = {Linux Foundation AI \& Data},
}

@article{boncz2020fsst,
  title={FSST: fast random access string compression},
  author={Boncz, Peter and Neumann, Thomas and Leis, Viktor},
  journal={Proceedings of the VLDB Endowment},
  volume={13},
  number={12},
  pages={2649--2661},
  year={2020},
  publisher={VLDB Endowment}
}

@article{lemire2015decoding,
  title={Decoding billions of integers per second through vectorization},
  author={Lemire, Daniel and Boytsov, Leonid},
  journal={Software: Practice and Experience},
  volume={45},
  number={1},
  pages={1--29},
  year={2015},
  publisher={Wiley Online Library}
}

@article{melnik2020dremel,
  title={Dremel: A decade of interactive SQL analysis at web scale},
  author={Melnik, Sergey and Gubarev, Andrey and Long, Jing Jing and Romer, Geoffrey and Shivakumar, Shiva and Tolton, Matt and Vassilakis, Theo and Ahmadi, Hossein and Delorey, Dan and Min, Slava and others},
  journal={Proceedings of the VLDB Endowment},
  volume={13},
  number={12},
  pages={3461--3472},
  year={2020},
  publisher={VLDB Endowment}
}

@article{melnik2010dremel,
  title={Dremel: interactive analysis of web-scale datasets},
  author={Melnik, Sergey and Gubarev, Andrey and Long, Jing Jing and Romer, Geoffrey and Shivakumar, Shiva and Tolton, Matt and Vassilakis, Theo},
  journal={Proceedings of the VLDB Endowment},
  volume={3},
  number={1-2},
  pages={330--339},
  year={2010},
  publisher={VLDB Endowment}
}

@article{bajaj2016ms,
  title={{MS MARCO}: A human generated machine reading comprehension dataset},
  author={Bajaj, Payal and Campos, Daniel and Craswell, Nick and Deng, Li and Gao, Jianfeng and Liu, Xiaodong and Majumder, Rangan and McNamara, Andrew and Mitra, Bhaskar and Nguyen, Tri and others},
  journal={arXiv preprint arXiv:1611.09268},
  year={2016}
}

@misc{bge-m3,
      title={BGE M3-Embedding: Multi-Lingual, Multi-Functionality, Multi-Granularity Text Embeddings Through Self-Knowledge Distillation}, 
      author={Jianlv Chen and Shitao Xiao and Peitian Zhang and Kun Luo and Defu Lian and Zheng Liu},
      year={2024},
      eprint={2402.03216},
      archivePrefix={arXiv},
      primaryClass={cs.CL}
}

@misc{cohere_wikipedia_22_12,
  author       = {{Cohere Team}},
  title        = {Wikipedia Dataset (December 2022)},
  year         = {2022},
  howpublished = {\url{https://huggingface.co/datasets/Cohere/wikipedia-22-12}},
  note         = {Accessed: 2025-10-27}
}

@inproceedings{wang2021milvus,
  title={Milvus: A purpose-built vector data management system},
  author={Wang, Jianguo and Yi, Xiaomeng and Guo, Rentong and Jin, Hai and Xu, Peng and Li, Shengjun and Wang, Xiangyu and Guo, Xiangzhou and Li, Chengming and Xu, Xiaohai and others},
  booktitle={Proceedings of the 2021 international conference on management of data},
  pages={2614--2627},
  year={2021}
}

@misc{pgvector,
  author       = {{pgvector Team}},
  title        = {pgvector: Open-source vector similarity search for PostgreSQL},
  year         = {2023},
  howpublished = {\url{https://github.com/pgvector/pgvector}},
  note         = {Accessed: 2025-10-27}
}

@misc{allenai_c4,
  author       = {{Allen Institute for AI}},
  title        = {C4 Dataset: Colossal Clean Crawled Corpus},
  year         = {2020},
  howpublished = {\url{https://huggingface.co/datasets/allenai/c4}},
  note         = {Accessed: 2025-10-27}
}

@inproceedings{gruenheid2025autocomp,
  title={{AutoComp}: Automated Data Compaction for Log-Structured Tables in Data Lakes},
  author={Gruenheid, Anja and Camacho-Rodr{\'\i}guez, Jes{\'u}s and Curino, Carlo and Ramakrishnan, Raghu and Pak, Stanislav and Sakdeo, Sumedh and Gandhi, Lenisha and Singhal, Sandeep K and Nilangekar, Pooja and Abadi, Daniel J},
  booktitle={Companion of the 2025 International Conference on Management of Data},
  pages={404--417},
  year={2025}
}

@article{agiwal2021napa,
  title={Napa: Powering scalable data warehousing with robust query performance at Google},
  author={Agiwal, Ankur and Lai, Kevin and Manoharan, Gokul Nath Babu and Roy, Indrajit and Sankaranarayanan, Jagan and Zhang, Hao and Zou, Tao and Chen, Min and Chen, Zongchang and Dai, Ming and others},
  journal={Proceedings of the VLDB Endowment},
  volume={14},
  number={12},
  pages={2986--2997},
  year={2021},
  publisher={VLDB Endowment}
}

@misc{VectorDBBench,
  title        = {VectorDBBench: A Benchmark Suite for Vector Databases},
  author       = {Zilliz},
  howpublished = {\url{https://github.com/zilliztech/VectorDBBench}},
  year         = {2023},
  note         = {Accessed: 2025-11-17}
}

@article{LambFVTVDB12,
  author       = {Andrew Lamb and
                  Matt Fuller and
                  Ramakrishna Varadarajan and
                  Nga Tran and
                  Ben Vandiver and
                  Lyric Doshi and
                  Chuck Bear},
  title        = {The Vertica Analytic Database: C-Store 7 Years Later},
  journal      = {Proc. {VLDB} Endow.},
  volume       = {5},
  number       = {12},
  pages        = {1790--1801},
  year         = {2012},
  url          = {http://vldb.org/pvldb/vol5/p1790\_andrewlamb\_vldb2012.pdf},
  doi          = {10.14778/2367502.2367518},
  bibsource    = {dblp computer science bibliography, https://dblp.org}
}

@inproceedings{prammer2025towards,
  title={Towards Functional Decomposition of Storage Formats},
  author={Prammer, Martin and Zeng, Xinyu and Meng, Ruijun and McKinney, Wes and Zhang, Huanchen and Pavlo, Andrew and Patel, Jignesh M},
  booktitle={Conference on Innovative Data Systems Research (CIDR)},
  year={2025}
}

\end{document}